\documentclass[preprint2]{proto}
\usepackage{natbib}
\bibpunct{(}{)}{;}{a}{,}{,}
\usepackage{times}
\makeatletter
\renewcommand{\subsubsection}{\@startsection
   {subsubsection}%
   {3}%
   {0em}%
   {-0.01\baselineskip}%
   {-2\fontdimen2\font 
      plus -2\fontdimen3\font
      minus -2\fontdimen4\font}
   {\normalfont\normalsize\itshape}}
\makeatother
\def\revised#1{{#1}}
\def\revrem#1{{}}
\def\remark#1{{}}
\def\MEMORY#1{{}}
\def\ppvcit#1{{#1}}
\def\sssec#1{\subsubsection{#1}}
\def\revadd#1{{#1}}
\def\removadd#1{}

\begin{document}

\title{\textbf{\LARGE Models of the Structure and Evolution of
    Protoplanetary Disks}}

\author {\textbf{\large C.P.~Dullemond}}
\affil{\small\em Max-Planck-Institute for Astronomy, Heidelberg}

\author {\textbf{\large D.~Hollenbach}}
\affil{\small\em NASA-Ames Research Center}

\author {\textbf{\large I.~Kamp}}
\affil{\small\em Space Telescope Division of ESA}

\author {\textbf{\large P.~D'Alessio}}
\affil{\small\em Centro de Radioastronom\'\i a y 
Astrof\'\i sica, UNAM, M\'exico}

\begin{abstract}
\baselineskip = 11pt
\leftskip = 0.65in
\rightskip = 0.65in
\parindent=1pc
{\small 
\noindent \revadd{We review advances in the modeling of protoplanetary
disks. This review will focus on the regions of the disk beyond the dust
sublimation radius, i.e.~beyond 0.1 - 1 AU, depending on the stellar
luminosity. We will be mostly concerned with models that aim to fit spectra
of the dust continuum or gas lines, and derive physical parameters from
these fits. For optically thick disks, these parameters include the
accretion rate through the disk onto the star, the geometry of the disk, the
dust properties, the surface chemistry and the thermal balance of the
gas. For the latter we are mostly concerned with the upper layers of the
disk, where the gas and dust temperature decouple and a photoevaporative
flow may originate.  We also briefly discuss
optically thin disks, focusing mainly on the gas, not the dust. The
evolution of these disks is dominated by accretion, viscous spreading,
photoevaporation, and dust settling and coagulation. The density and
temperature structure arising from the surface layer models provide input to
models of photoevaporation, which occurs largely in the outer disk. We
discuss the consequences of photoevaporation on disk evolution and planet
formation.}  \\~\\~\\~}
 
\end{abstract}

\section{\textbf{INTRODUCTION}}
\label{sec-intro}\noindent
Dusty circumstellar disks have been the focus of intense observational
interest in recent years, largely because they are thought to be the
birthplaces of planetary systems. These observational efforts have yielded
many new insights on the structure and evolution of these disks. In spite of
major developments in spatially resolved observations of these disks, much
of our knowledge of their structure is still derived from spatially {\em
un}resolved spectroscopy and spectral energy distributions (SEDs). The
interpretation of this information (as well as spatially resolved data)
requires the use of theoretical models, preferentially with as much realism
and self-consistency as possible. Such disk models have been developed and
improved over many years.  When they are in reasonable agreement with
observations they can also serve as a background onto which other processes
are modeled, such as chemistry, grain growth, and ultimately the formation
of planets.

This chapter reviews the development of such self-consistent disk structure
models, and discusses the current status of the field. \revised{We focus on
the regions of the disk beyond the dust sublimation radius, since the very
inner regions are discussed in the \ppvcit{chapter by Najita et al.}. 
To limit our
scope further,} we restrict our review to models primarily aimed at a
comparison with observations. We will start with a concise resum\'e of the
formation and viscous evolution of disks (Section \ref{sec-viscevol}). This
sets the radial disk structure as a function of time. We then turn our
attention to the vertical structure, under the simplifying assumption that
the gas temperature equals the dust temperature everywhere (Section
\ref{sec-diskstruct}). While this assumption is valid in the main body of
the disk, it breaks down in the disk surface layers. \revised{Section
\ref{sec-surface} treats the gas physics and chemistry of these surface
layers, where much of the spectra originate. Photoevaporation flows also
originate from the warm surface layers, and affect the disk evolution, which
is the topic of Section \ref{sec-evaporation}.}

\section{\textbf{FORMATION AND VISCOUS EVOLUTION OF DISKS}}
\label{sec-viscevol}
\begin{figure}[tb]
\centerline{\includegraphics[width=8cm]{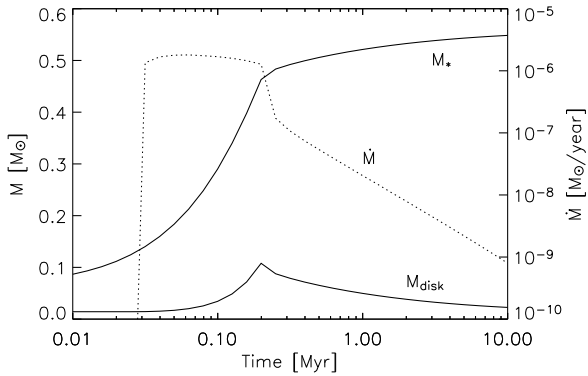}}
\caption{\label{fig-hueso}
Evolution of various disk and star quantities as a function of time after
the onset of collapse of the cloud core (after {\it Hueso and
Guillot}, \citeyear{huesoguillot:2005}). Solid lines: stellar mass (upper)
and disk mass (lower). Dotted line: accretion rate {\em in the disk}.
\revadd{In this model the disk is formed at $t\simeq 0.03$ Myr, causing the
jump in the dotted line at this point. The collapse phase is
finished by $2\times 10^5$ years.}}
\end{figure}

\noindent
The formation of stars and planetary systems starts with the gravitational
collapse of a dense molecular cloud core. Since such a core will always have
some angular momentum at the onset of collapse, most of the infalling matter
will not fall directly onto the protostar, but form a disk around it
\revised{(e.g.,~{\it Terebey et al.}, \citeyear{terebyshucas:1984}; {\it
Yorke et al.}, \citeyear{yorkebodlau:1993}) or fragment into a multiple
stellar system (e.g.,~{\it Matsumoto and
Hanawa}, \citeyear{matsuhana:2003}). Because of the complexity of the
latter, we focus on the single star formation scenario here.}  While matter
falls onto the disk, viscous stresses \revised{and gravitational torques}
within the disk will transport angular momentum to its outer regions. As a
consequence of this, most of the disk matter moves inward, adding matter to
the protostar, while some disk matter moves outward, absorbing all the
angular momentum ({\it Lynden-Bell and
Pringle}, \citeyear{1974MNRAS.168..603L}). During its formation and
evolution a disk will spread out to several 100 AU or more ({\it Nakamoto
and Nakagawa}, \citeyear{nakamotonakagawa:1994}, henceforth NN94; {\it Hueso
and Guillot}, \citeyear{huesoguillot:2005}, henceforth HG05). This spreading
is only stopped when processes such as photoevaporation (this chapter),
stellar encounters ({\it Scally and Clarke}, \citeyear{scallyclarke:2001};
{\it Pfalzner et al.}, \citeyear{pfalzner:2005}) or a binary companion ({\it
Artymowicz and Lubow}, \citeyear{artymowicz:1994}) truncate the disk from
the outside. During the collapse phase, which lasts a few$\times 10^5$
years, the accretion rate within the disk is very high ($\dot M\sim
10^{-5}$ -- $10^{-6}\,M_{\odot}/$yr), but quickly drops to $\dot M\sim
10^{-7}$ -- $10^{-9}\,M_{\odot}/$yr once the infall phase is over (NN94,
HG05). The optical and ultraviolet excess observed from classical T Tauri
stars (CTTSs) and Herbig Ae/Be stars (HAeBes) confirms that this on-going
accretion indeed takes place ({\it Calvet et
al.}, \citeyear{calvethartstrom:2000}, and references therein). In
Fig.~\ref{fig-hueso} we show the evolution of various disk and star
parameters.

\subsection{Anomalous viscosity}
An issue that is still a matter of debate is what constitutes the viscosity
required for disk accretion, \revised{particularly after infall has ceased.}
Molecular viscosity is too small to account for the observed mass accretion
rates. Turbulent and magnetic stresses, however, can constitute some kind of
anomalous viscosity. The magnetorotational instability (MRI), present in
weakly magnetized disks, is the most accepted mechanism to drive turbulence
in disks and transport angular momentum outwards ({\it Balbus and
Hawley}, \citeyear{balbushawley:1991}; {\it Stone and
Pringle}, \citeyear{stonepringle:2001}; {\it
Wardle}, \citeyear{wardle:2004}, and references therein).

There is a disk region (0.2 $< r <$ 4 AU, for typical CTTS disk parameters
according to {\it D'Alessio et al.}, \citeyear{dalessiocanto:1998}) in which
the ionization fraction is smaller than the minimum value required for the
MRI. Neither thermal ionization (requiring a temperature higher than 1000
K), cosmic ray ionization (requiring a mass surface density smaller than
$\sim$~100 g/cm$^2$) ({\it Jin}, \citeyear{jin:1996}; {\it
Gammie}, \citeyear{gammie:1996}), nor X-rays ({\it Glassgold et
al.}, \citeyear{glassgoldnajita:1997}, \citeyear{glassgoldnajita:1997err})
are able to provide a sufficient number of free electrons to have MRI
operating near the midplane.  {\it Gammie}~(\citeyear{gammie:1996}) proposed
a layered accretion disk model, in which a ``dead zone'' is encased between
two actively accreting layers. The precise extent of this dead zone is
difficult to assess, because the number density of free electrons depends on
detailed chemistry as well as the dust grain size distribution, since dust
grains tend to capture free electrons ({\it Sano et
al.}, \citeyear{sanomirama:2000}, and references therein). If the disk dust
is like the interstellar dust, the MRI should be inhibited 
in large parts of the disk 
({\it Ilgner and Nelson}, \citeyear{ilgnernelson:06a}), 
though this is still under debate (e.g.~{\it Semenov et
al.}, \citeyear{semenovwiebe:2004}; \ppvcit{see chapter by Bergin et al.}).

There are also other (non-magnetic) mechanisms for anomalous viscosity, like
the baroclinic instability ({\it Klahr and
Bodenheimer}, \citeyear{klahrboden:2003}) or the shear instability ({\it
Dubrulle et al.}, \citeyear{dubrullemarie:2005}), which are still subject to
some controversy (see the recent review by {\it Gammie and
Johnson}, \citeyear{gammiejohnson:2005}). Angular momentum can also be
transferred by global torques, such as through gravitational spiral waves
({\it Tohline and Hachisu}, \citeyear{tohlinehachisu:1990}; {\it Laughlin and
Bodenheimer}, \citeyear{1994ApJ...436..335L}; {\it Pickett et
al.}, \citeyear{pickett:2003} and references therein) 
or via global magnetic fields threading the disk ({\it Stehle
and Spruit}, \citeyear{stehlespruit:2001}), possibly with hydromagnetic
winds launched along them ({\it Blandford and
Payne}, \citeyear{blandfordpayne:1982}; {\it Reyes-Ruiz and
Stepinski}, \citeyear{reyes-ruiz:1996}).

\subsection{$\alpha$-Disk models for protoplanetary disks}
To avoid having to solve the problem of viscosity in detail, but still be
able to produce sensible disk models, {\it Shakura and
Sunyaev}~(\citeyear{shaksuny:1973}) introduced the
``$\alpha$-prescription'', based on dimensional arguments. In this recipe
the vertically averaged viscosity $\nu$ at radius $r$ is written as
$\nu=\alpha H_p c_s$, where $H_p$ is the pressure scale height of the disk
and $c_s$ is the isothermal sound speed, both evaluated at the disk midplane
where most of the mass is concentrated. The parameter $\alpha$ summarizes
the uncertainties related to the sources of anomalous viscosity, and is
often taken to be of the order of $\alpha\simeq 10^{-2}$ for sufficiently
ionized disks.

From conservation of angular momentum, the mass surface density $\Sigma$ of
a {\em steady} disk (i.e.~with a constant mass accretion rate $\dot{M}$),
for radii much larger than the disk inner radius $r_{\mathrm{in}}$, can be
written as $\Sigma(r) \approx \dot{M} / 3 \pi \nu$. With $H_p=c_s/\Omega_K$,
where $\Omega_K$ is the Keplerian angular velocity, we see that for $r \gg
r_{\mathrm{in}}$,
\begin{equation}\label{eq-sigma-from-visc}
\Sigma(r) = K \;
\frac{\dot{M}}{r^{3/2}\alpha T_c(r)}
\;,
\end{equation}
\revised{\revadd{where $T_c(r)$ is the midplane temperature of the disk at
radius $r$ and $K$ is a constant with the value $K\equiv \sqrt{GM_{*}}\mu
m_p/3\pi k$. Here $\mu$ is the mean molecular weight in units of the proton
mass $m_p$, $G$ is the gravitational constant, $k$ is Boltzmann's constant
and $M_{*}$ is the stellar mass.}  As we will show in Section
\ref{sec-diskstruct}, most of the disk is `irradiation-dominated', and
consequently has temperature given approximately by $T_c\sim r^{-1/2}$. This
results in the surface density going as $\Sigma\sim r^{-1}$. This surface
density distribution is less steep than the so-called ``minimum mass solar
nebula'' (MMSN), given by $\Sigma \sim r^{-3/2}$ ({\it
Weidenschilling}, \citeyear{weidenmmsn:1977}; {\it
Hayashi}, \citeyear{hayashi:1981}). Strictly, the MMSN does not necessarily
represent the mass distribution at any instant, but the minimum mass that
has passed through the disk during its lifetime ({\it
Lissauer}, \citeyear{lissauer:1993}, and references therein).}

\revised{In reality protoplanetary disks are not quite steady. After the
main infall phase is over, the disk is not supplied anymore with new matter,
and the continuing accretion onto the star will drain matter from the disk
(see Fig.~\ref{fig-hueso}). In addition the disk viscously expands and is
subject to photoevaporation (see Section \ref{sec-evaporation}). The
timescale for ``viscous evolution'' depends on radius and is given by
$t_{\mathrm{vis}} \simeq r^2/\nu$, which for typical CTTS parameters is 1
Myr at $r\simeq$ 100 AU. Since for irradiated disks $t_{\mathrm{vis}} \propto
r$, the outer regions evolve the slowest, yet they contain most of the
mass. These regions ($\gtrsim 50-100$ AU) therefore form a reservoir of mass
constantly resupplying the inner regions. The latter can thus be approximately
described by steady accretion disk models.}

\revised{There might also be dramatic variability taking place on shorter
timescales,} as shown by FU Ori and EX Lupi type outbursts ({\it Gammie and
Johnson}, \citeyear{gammiejohnson:2005}, and references therein). These
outbursts can have various triggering mechanisms, such as thermal
instability ({\it Kawazoe and Mineshige}, \citeyear{kawazoe:1993}; {\it Bell
and Lin}, \citeyear{belllin:1994}); close passage of a companion star ({\it
Bonnell and Bastien} \citeyear{bonnelbastien:1992}; {\it Clarke and
Syer}, \citeyear{clarkesyer:1996}); mass accumulation in the dead zone
followed by gravitational instability ({\it Gammie}, \citeyear{gammie:1996};
{\it Armitage et al.}, \citeyear{2001MNRAS.324..705A}). Disks are therefore
quite time-varying, and constant $\alpha$ steady disk models should be taken
as zeroth-order estimates of the disk structure.

Given the challenges of understanding the disk viscosity from first
principles, attempts have been made to find observational constraints on
disk evolution ({\it Ruden and Pollack}, \citeyear{rudenpollack:1991}; {\it
Cassen}, \citeyear{cassen:1996}; {\it Hartmann et
al.}, \citeyear{hartcalvgulldal:1998}; {\it
Stepinski}, \citeyear{stepinski:1998}). For example, {\it Hartmann et
al.}~(\citeyear{hartcalvgulldal:1998}) study a large sample of CTTSs and
find a decline in mass accretion rate with time, roughly described as
$\dot{M} \sim t^{-1.5}$, which they compare to the analytic similarity
solutions of {\it Lynden-Bell and Pringle}~(\citeyear{1974MNRAS.168..603L})
for the expanding disk. \revised{A similar type of observational constraint
is the recently found rough correlation $\dot M\propto M_{*}^2$ ({\it
Muzerolle et al.}, \citeyear{muzerollehillen:2003}, \citeyear{muzerolle:2005}; 
{\it Natta et al.}, \citeyear{nattatesti:2004}).} 
\revised{High angular resolution mm-wave
continuum imaging can also help to constrain the mass surface density
distribution. With this technique {\it Wilner et
al.}~(\citeyear{wilnerhokast:2000}) concluded that $\Sigma\propto r^{-1}$
for TW Hydra. {\it Kitamura et al.}~(\citeyear{kitamuramomose:2002}) find
$\Sigma \sim r^{-p}$, with $p=0-1$ for a sample of T Tauri stars.}

\section{\textbf{VERTICAL STRUCTURE OF DUSTY DISKS}}
\label{sec-diskstruct}\noindent
With the radial structure following from accretion physics, as described
above, the next issue is the vertical structure of these disks.  Many
authors have modeled this with full time-dependent 2D/3D
(magneto/radiation-) hydrodynamics (e.g.,~{\it
Boss}, \citeyear{boss:1996}, \citeyear{boss:1997err}; {\it Yorke and
Bodenheimer}, \citeyear{yorkeboden:1999}; {\it Fromang et
al.}, \citeyear{fromangbalbusdevil:2004}). While this approach is obviously
\revised{very powerful}, it suffers from large computational costs, and
often requires strong simplifying assumptions in the radiative transfer to
keep the problem tractable.  For comparison to \revised{observed spectra and
images} these models are therefore less practical.  \revised{Most
observation-oriented disk structure models split the disk into a series of
(nearly independent) annuli, each constituting a 1-D or a two-layer local
vertical structure problem.  In this section we review this kind of `1+1D'
models, and their 2-D/3-D generalizations.}

%
\subsection{Basic principles}\noindent
The main objective of the models described in this section is the
determination of the density and temperature structure of the disk. For a
given surface density $\Sigma(r)$, and a given \revised{gas} temperature
structure $T_g(r,z)$ (where $z$ is the vertical coordinate measured upward
from the midplane) the vertical density distribution $\rho(r,z)$ can be
readily obtained by integrating the vertical equation of hydrostatics:
\begin{equation}
\frac{dP}{dz}=-\rho\,\Omega_K^2\;z
\end{equation}
where $P=\rho c_s^2$ with $c_s^2\equiv k\,T_g/\mu\,m_p$. \revised{The main
complexity of a disk model lies in the computation of the {\em temperature}
structure.}  Since the main source of opacity is the dust, most models so
far make the assumption that the gas temperature is equal to the dust
temperature, \revised{so that the gas temperature determination reduces
to solving a dust continuum radiative transfer problem.}  \revised{In
Section \ref{sec-surface} we will relax this assumption, but until then we
will keep it.}

The temperature of the disk is set by a balance between heating and
cooling. The disk cools by thermal emission from the dust grains
at infrared wavelengths. This radiation is what is observed as infrared dust
continuum radiation from such disks. Line cooling is only a minor coolant,
and only plays a role for $T_g$ when gas and dust are thermally decoupled.
Dust grains can be heated in part by radiation from other grains in the
disk. The iterative absorption and re-emission of infrared radiation by dust
grains in the disk causes the radiation to propagate \revised{through} the
disk in a diffusive way. \revised{Net energy input comes from absorption of
direct stellar light in the disk's surface layers, and from} viscous
dissipation of gravitational energy in the disk due to
accretion. \revised{For most disks around CTTSs and Herbig Ae/Be stars the
heating by stellar radiation is dominant over the viscous heating (except in
the very inner regions). Only for strongly accreting disks 
does the latter dominate.}

\begin{figure}[tb]
\centerline{\includegraphics[width=8cm]{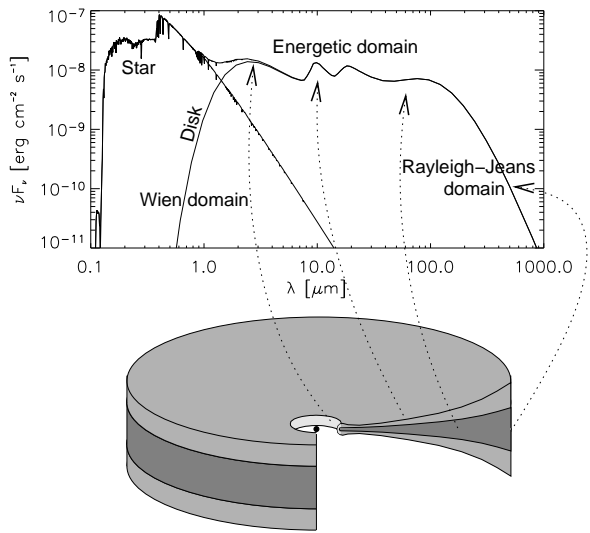}}
\caption{\label{fig-sed-and-disk}\revised{Build-up of the SED of a flaring
protoplanetary disk and the origin of various components: the near-infrared
bump comes from the inner rim, the infrared dust features from the warm
surface layer, and the underlying continuum from the deeper (cooler) disk
regions. Typically the near- and mid-infrared emission comes from small
radii, while the far-infrared comes from the outer disk regions. The
(sub-)millimeter emission mostly comes from the midplane of the outer
disk. Scattering is not included here.}}
\end{figure}

Once the temperature structure is determined, the SED can be computed. The
observable thermal emission of a dusty disk model consists of three
wavelength regions \revised{(see Fig.~\ref{fig-sed-and-disk})}. The main
portion of the energy is emitted in a wavelength range depending on the
minimum and maximum temperature of the dust in the disk. We call this the
``energetic domain'' of the SED, which typically ranges from 1.5 $\mu$m to
about 100 $\mu$m. At shorter wavelength the SED turns over into the ``Wien
domain''.  At longer wavelengths the SED turns over into the "Rayleigh-Jeans
domain", a steep, nearly powerlaw profile with a slope depending on grain
properties and disk optical depth (\ppvcit{see chapter by Natta et al.}).
\revised{Differences in disk geometry are mainly reflected in the energetic
domain of the SED, while the submm and mm fluxes probe the disk mass.}

%
\subsection{A first confrontation with observations}\noindent
\revised{It is quite challenging to solve the entire disk structure
according to the above principles. Early disk models were therefore often
based on strong simplifications.} \revised{An example of such a model} is a
perfectly flat disk being irradiated by the star due to the star's
non-negligible size ({\it Adams and Shu}, \citeyear{adamsshu:1986}; {\it
Friedjung}, \citeyear{friedjung:1985}).  The stellar radiation impinges onto
the flat disk under an irradiation angle $\varphi\simeq 0.4 r_{*}/r$ (with
$r_{*}$ the stellar radius). Neglecting viscous dissipation, the effective
temperature of the disk is set by a balance between the irradiated flux
\revised{$(1/2)\varphi L_{*}/4\pi r^2$} (with $L_{*}$ the stellar
luminosity) and blackbody cooling
\revised{$\sigma T_{\mathrm{eff}}^4$}, which yields $T_{\mathrm{eff}}\propto
r^{-3/4}$. The energetic domain of its SED therefore has a slope of $\nu
F_\nu\propto \nu^{s}$ with $s=4/3=1.33$, \revised{which} follows from the
\revised{fact} that any disk with $T_{\mathrm{eff}}\propto r^{-q}$ has an
SED slope of $s=(4q-2)/q$.  \revised{This steep slope arises because most of
the stellar radiation is absorbed and re-emitted at small radii where the
disk is hot. This produces strong emission at short wavelength. The long
wavelength flux is weak because only little stellar radiation is absorbed at
large radii.}  Observations of CTTSs, however, show SED slopes typically in
the range $s=0.6$ to 1 ({\it Kenyon and
Hartmann}, \citeyear{kenyonhart:1995}), i.e.~much less steep. The SEDs of
Herbig Ae/Be stars show a similar picture, \revised{but} with a somewhat
larger spread in $s$\revised{, though it must be kept in mind that the
determination of the slope of a bumpy SED like in
Fig.~\ref{fig-sed-and-disk} is somewhat subjective}.  {\it Meeus et
al.}~(\citeyear{meeuswatersbouw:2001}, henceforth M01) divide the SEDs of
Herbig Ae/Be stars into two groups: those with strong far-infrared flux
(called `group I', having slope $s\simeq -1$ ... $0.2$) and those with weak
far-infrared flux (called `group II', having slope $s\simeq 0.2$ ... $1$). All
but the most extreme group II sources have a slope that is clearly
inconsistent with that of a flat disk. \revised{Note, at this point, that
the Meeus `group I' and `group II' are unrelated to the Lada `class I' and
`class II' classification (both Meeus group I and II are members of Lada
class II).}

\revised{A number of authors have employed another model to interpret their
observations of protoplanetary disks: that of a steady accretion disk heated
by viscous dissipation ({\it Rucinski et al.}, \citeyear{rucinski:1985};
\revised{{\it Bertout et al.}, \citeyear{bertoutbasribouv:1988};} {\it
Hillenbrand et al.}, \citeyear{hillenstrom:1992}). These models are based on
the model by {\it Shakura and Sunyaev}~(\citeyear{shaksuny:1973}).
\revadd{A detailed vertical structure model of such a disk was presented by
{\it Bell et al.}~(\citeyear{bellcassklhen:1997}).} \revadd{The luminosity
of such disks, including the magnetospheric accretion column, is 
$L_{\mathrm{accr}}=G M_{*}\dot M/r_{*}$.} \revadd{For $r\gg r_{\mathrm{in}}$
the effective temperature of such disks is given by $\sigma
T_{\mathrm{eff}}^4=3\dot M\Omega_K^2/8\pi$ (with $\sigma$ the
Stefan-Boltzmann constant), yielding an SED slope of $s=4/3$, like for
passive flat disks ({\it Lynden-Bell}, \citeyear{lynden-bell:1969}; see
solid lines of Fig.~\ref{fig-seds-accr}).} Therefore these models are not
very succesful either, except for modeling very active disks like FU Orionis
(FUor) outbursts (see {\it Bell and Lin}, \citeyear{belllin:1994}).}

%
\subsection{Flaring disk geometry}\noindent
It was recognized by {\it Kenyon and Hartmann}~(\citeyear{kenyonhart:1987}) 
that a natural explanation for the strong far-infrared flux
(i.e.~shallow SED slope) of most sources is a flaring
\revised{(``bowl-shaped'')} geometry of the disk\revised{'s surface, as
depicted in Fig.~\ref{fig-sed-and-disk}}.  The flaring geometry
\revised{allows the disk to capture} a significant portion of the stellar
radiation at large radii where the disk is cool, \revised{thereby boosting
the mid- to far-infrared emission.}

\revised{The flaring geometry adds an extra term to the irradiation angle:
$\varphi\simeq 0.4\;r_{*}/r+\,rd(H_s/r)/dr$ ({\it Chiang and
Goldreich}, \citeyear{chianggold:1997}, henceforth CG97),} where $H_s$ is
the height above the midplane where the disk becomes optically thick to the
impinging stellar radiation. \revised{In the same way as for the flat disks
the thermal balance determines the $T_{\mathrm{eff}}$ of the disk, but this
now depends strongly on the shape of the disk: $H_s(r)$. The pressure scale
height $H_p$, on the other hand, depends on the midplane temperature $T_c$
by $H_p=\sqrt{k T_{\mathrm{c}} r^3/\mu m_p G M_{*}}$ (with $M_{*}$ the
stellar mass). If we set $T_c=T_{\mathrm{eff}}$ and if the ratio $\chi\equiv
H_s/H_p$ is known, then the system of equations is closed and can be solved
(see appendix {\it Chiang et al.},~{\citeyear{chiangjoung:2001}}). For the
special case that $\chi$ is constant we obtain $H_s\propto r^{9/7}$,
a-posteriori confirming that the disk indeed has a ``bowl'' shape. In
general, though, $\chi$ must be computed numerically, and depends on the
dust opacity of the disk upper layers. The resulting temperature profile
is typically about $T_{\mathrm{c}}\propto r^{-0.5}$.}

\revadd{The total luminosity of such a non-accreting flaring disk is
$L_{\mathrm{disk}}=C\,L_{*}$, where $C$ is the {\em covering fraction} of
the disk. The covering fraction is the fraction of the starlight that is
captured by the material in the disk. With a large enough disk inner radius
($r_{\mathrm{in}}\gg r_{*}$) one can write $C\simeq \max(H_s(r)/r)$.  For an
infinitely thin disk extending from $r_{*}$ to $r\rightarrow \infty$ one has
$C=0.25$. The {\em observed} flux
ratio $F_{\mathrm{disk}}/F_{*}$ may deviate from $C$ by about a factor of
2 due to the anisotropy of disk emission.}

\revised{In addition to irradiation by the star, the outer regions of a
strongly accreting flaring disk (like an FUor object) can also be irradiated
by the accretion luminosity from the inner disk ({\it Kenyon and
Hartmann}, \citeyear{kenyonhart:1991}; {\it Bell}, \citeyear{bell:1999};
{\it Lachaume}, \citeyear{lachaume:2004}) \revadd{and by the emission from
the magnetospheric accretion column or boundary layer ({\it Muzerolle et
al.}, \citeyear{muzerollecalvet:2003}).}}

\subsection{Warm dust surface layer}\noindent
A closer look at the physics of an irradiation-dominated disk (be it flat or
flared) reveals that its surface temperature is generally higher than its
interior temperature ({\it Calvet et al.}, \citeyear{calvetpatino:1991};
{\it Malbet and Bertout}, \citeyear{malbetbertout:1991};
\revadd{CG97}). \revised{Dust grains in the surface layers are directly
exposed to the stellar radiation, and are therefore hotter than dust grains
residing deep in the disk which only `see' the infrared emission by other
dust grains. The temperature difference is typically a factor of $2$ -- $4$
for non/weakly-accreting disks (see curve labeled ``-9'' in
Fig.~\ref{fig-vstruct}). For non-negligible accretion, on the other hand,
the disk is heated from inside as well, producing a temperature minimum
somewhere between the equatorial plane and the surface layer (see other
curves in Fig.~\ref{fig-vstruct}).}  Because of the shallow incidence angle
of the stellar radiation $\varphi\ll 1$, the {\em vertical} optical depth of
this warm surface layer is very low.  \revadd{The layer produces optically
thin emission at a temperature higher than the effective temperature of the
disk. The frequency-integrated flux of this emission is the same as that
from the disk interior.} As a consequence, the thermal radiation from these
surface layers produces dust features in {\em emission}. This is exactly
what is seen in nearly all non-edge-on T Tauri and Herbig Ae/Be star spectra
(e.g.\ M01; {\it Kessler-Silacci et al.}, \citeyear{kessler:2006}), indicating
that these disks are nearly always dominated by irradiation.

\subsection{Detailed models for flaring disks}\noindent
\sssec{Disk structure.}\noindent
Armed with the concepts of disk flaring and hot surface layers, a number of
authors published detailed \revised{1+1D disk models and two-layer
(surface+interior) models} with direct applicability to observations. The
aforementioned CG97 model (with refinements described in {\it Chiang et
al.}, \citeyear{chiangjoung:2001}) is a \revadd{handy} two-layer model for
the interpretation of SEDs and dust emission features from {\em
non-accreting} (`passive') disks. 
{\it Lachaume et al.}~(\citeyear{lachaume:2003}) extended it to include
viscous dissipation.

\begin{figure}[tb]
\centerline{\includegraphics[width=8cm]{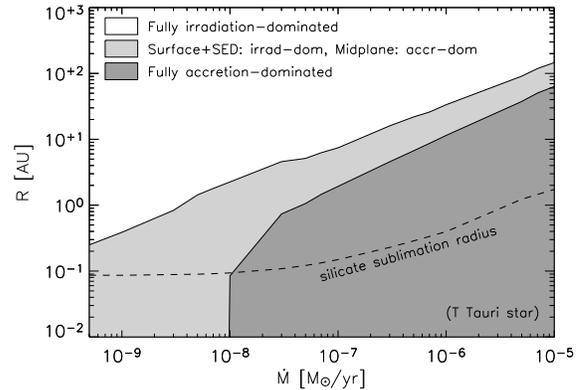}}
\caption{\label{fig-zones}
\revadd{Characteristic radii of a disk around a $0.9\;M_{\odot}$ star with
$T_{*}=4000\;$K and $R_{*}=1.9\;R_{\odot}$ at different accretion rates.
The silicate sublimation radius is the dust `inner rim'. 
Figure based on models by 
{\it D'Alessio et al.}~(\citeyear{dalessiocanto:1998}).}
}
\end{figure}

\begin{figure}[tb]
\centerline{\includegraphics[width=8cm]{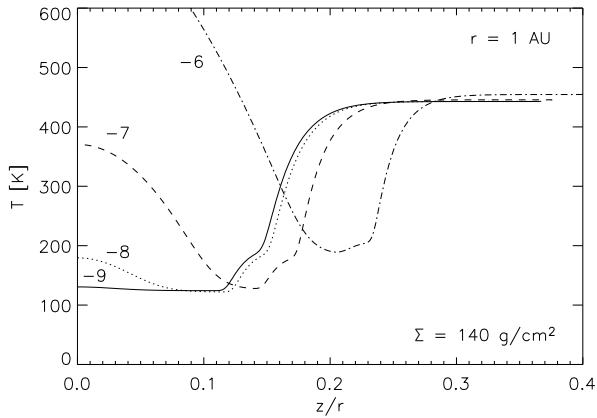}}
\caption{\label{fig-vstruct}
Vertical temperature distribution of an irradiated $\alpha$-disk at 1 AU,
for a fixed $\Sigma$ (chosen to be that of a disk model with $\dot M=10^{-8}
\,M_\odot/$yr for $\alpha=0.01$), but varying $\dot M$, computed using the
models of {\it D'Alessio et
al.}~(\citeyear{dalessiocanto:1998}). \revadd{The labels of the curves
denote the 10-log of the accretion rate in $M_{\odot}$/yr.}}
\end{figure}

The models by {\it D'Alessio et al.}~(\citeyear{dalessiocanto:1998}) solve
the complete 1+1D disk structure problem \revadd{with diffusive radiative
transfer,} including \revised{stellar} irradiation and viscous dissipation
(using the $\alpha$ prescription). The main input parameters are a global
(constant) mass accretion rate $\dot M$ and $\alpha$. The surface density
profile $\Sigma(r)$ is calculated self-consistently. \revadd{This model
shows that the disk can be divided into three zones: an outer zone in which
the disk is dominated by irradiation, an inner zone where viscous
dissipation dominates the energy balance, and an intermediate zone where the
midplane temperature is dominated by the viscous dissipation but the surface
temperature by irradiation (See Fig.~\ref{fig-zones}).  In the intermediate
zone the vertical disk thickness is set by $\dot{M}$ and $\alpha$ 
but the infrared spectrum is still powered
by irradiation. In Fig.~\ref{fig-vstruct} the vertical structure of the disk
is shown, for fixed $\Sigma$ but varying $\dot M$ for $r=1$ AU.}

\removadd{Removed the sentence about how the transition radii change with
$\dot M$: with the extra figure this is no longer necessary.}

\revadd{The models described by {\it Dullemond, et
al.}~(\citeyear{dulvzadnat:2002}) apply exact 1-D wavelength-dependent
radiative transfer for the vertical structure, but these models do not
include viscous dissipation.}

\vspace{1em}

\sssec{Dust growth and sedimentation.}
\noindent
Models of \revised{the kind discussed above} describe the SEDs of CTTSs
reasonably well. However, {\it D'Alessio et
al.}~(\citeyear{dalessiocalvet:1999}) argue that they tend to slightly
overproduce far-infrared flux and have too thick dark lanes in images of
edge-on disks. They also show that the percentage of expected edge-on disks
appears to be overpredicted. They suggest that dust sedimentation could help
to solve this problem. {\it Chiang et al.}~(\citeyear{chiangjoung:2001})
find similar results for a subset of their Herbig Ae/Be star sample: the
Meeus group
II sources (see also CG97). They fit these sources by \revised{mimicking} dust
settling through a reduction of the disk surface height.
Self-consistent computations of dust sedimentation produce similar SEDs and
confirm the dust settling idea ({\it Miyake and
Nakagawa}, \citeyear{miyakenaka:1995}; {\it Dullemond and
Dominik}, \citeyear{duldomsett:2004}, henceforth DD04b; {\it D'Alessio et
al.}, \citeyear{dalessio:06a}). The disk thickness and far-infrared flux can
also be reduced by grain growth ({\it D'Alessio et
al.}, \citeyear{dalessiocalvet:2001}; {\it Dullemond and
Dominik}, \citeyear{duldomdisk:2004}). \revised{The \ppvcit{chapter by
Dominik et al.\ }discusses such models of grain growth and sedimentation in
detail.}

From comparing infrared and (sub-)millimeter spectra of the same sources
({\it Acke et al.}, \citeyear{acke:2004a}), it is clear that \revised{small
and big grains co-exist in these disks.}  The (sub-)millimeter
\revised{spectral slopes} usually require mm-sized grains \revised{near the
midplane} in the outer regions of the disk, while infrared dust emission
features clearly prove that the disk surface layers are dominated by grains
no larger than a few microns (see \ppvcit{chapter by Natta et al.}). It
appears that a bimodal \revised{grain} size distribution can fit the
observed spectra: \revised{a portion of sub-micron grains in the surface
layers responsible for the infrared dust emission features and a portion of
mm-sized grains in the disk interior accounting for the (sub-)millimeter
emission ({\it Natta et al.}, \citeyear{nattaprusti:2001}).}

\subsection{The dust `inner rim'}
The very inner part of the disk is dust-free due to dust sublimation (see
\ppvcit{chapter by Najita et al.\ }for a discussion of this region).  The
dusty part of the disk can therefore be expected to have a relatively abrupt
inner edge at about $0.5$ AU for a 50$\,L_{\odot}$ star (scaling roughly with
$\sqrt{L_{*}}$). If the gas inward of this dust inner rim is optically thin,
which \revised{appears to be} mostly the case ({\it Muzerolle et
al.}, \citeyear{muzerolle:2004}), then this dust inner rim is illuminated by
the star at a $\sim\,$90 degree angle, and is hence expected to be much hotter than
the rest of the disk behind it \revised{which is irradiated under a shallow
angle $\varphi\ll 1$.} \revadd{({\it Natta et
al.}, \citeyear{nattaprusti:2001})}. \revised{Consequently it must be
hydrostatically `puffed-up', although this is still under
debate}. \revadd{{\it Natta et al.}~(\citeyear{nattaprusti:2001}) showed
that the emission from such a hot inner rim can explain the near-infrared
bump seen in almost all Herbig Ae/Be star SEDs (see e.g.,~M01).} This is a
natural explanation, since dust sublimation occurs typically around 1500 K,
and a 1500 K blackbody bump fits reasonably well to the near-infrared bumps
in those sources. 
{\it Tuthill et al.}~(\citeyear{tutmondan:2001})
independently discovered a bright half-moon ring around the Herbig Be star
LkHa-101, which they attribute to a bright inner disk rim due to dust
sublimation.  {\it Dullemond et al.}~(\citeyear{duldomnat:2001}; henceforth
DDN01) extended the CG97 model to include such a puffed-up rim, and {\it
Dominik et al.}~(\citeyear{domdulwatwal:2003}) showed that the Meeus sample
of Herbig Ae/Be stars can be reasonably well fitted by this model. However,
for Meeus group II sources these fits required relatively small disks
(see, however, Section~\ref{subsec-2dtrans}).

The initial rim models were rather simplified, treating it as a vertical
blackbody `wall' (DDN01). {\it Isella and Natta}~(\citeyear{isella:2005})
improved this by noting that the weak dependence of the
sublimation temperature on gas density is enough to strongly round off the
rim.
Rounded-off rims appear to be more consistent with observations than the
vertical ones: \revised{their flux is less inclination dependent, and their
images on the sky are not so much one-sided}. There is still a worry,
though, whether the rims can be high enough to fit sources with a 
strong near-infrared bump.

With near-infrared interferometry the rim can be spatially resolved, and
thus the models can be tested. The measurements so far do
not yet give images, but the measured `visibilities' can be compared to
models. In this way one can measure the radius of the rim (e.g.,~{\it Monnier
et al.}, \citeyear{monnier:2005}; {\it Akeson et al.}, \citeyear{akeson:2005})
and its inclination (e.g.,~{\it Eisner et
al.}, \citeyear{eislanake:2003}). Moreover it can test whether indeed the
near-infrared emission comes from the inner rim of the dust disk in the
first place (some doubts have been voiced by {\it Vinkovic et
al.}, \citeyear{vinkovic:2003}). We refer to the \ppvcit{chapter by
Millan-Gabet et al.\ }for a more in-depth discussion of interferometric
measurements of disks.

The inner rim model has so far been mainly applied to Herbig Ae/Be stars
because the rim appears so apparent in the \revadd{near-infrared} (NIR). But
{\it Muzerolle et al.}~(\citeyear{muzerollecalvet:2003}) showed that it
also applies to T Tauri stars. In that case, however, the luminosity from
the magnetospheric accretion shock is required in addition to the stellar
luminosity to power the inner rim emission.

\revadd{In addition to being a strong source of NIR flux, the
`puffed-up' inner dust rim might also be responsible for the irregular
few-day-long extinction events observed toward UX Orionis stars ({\it Natta
et al.}~\citeyear{nattaprusti:2001}; {\it Dullemond et
al.}~\citeyear{dulancackboe:2003}). The latter authors argued that
this only works for self-shadowed (or only weakly flaring) disks
(see Section\ \ref{subsec-2dtrans}).}

\begin{figure}[tb]
\centerline{\includegraphics[width=8cm]{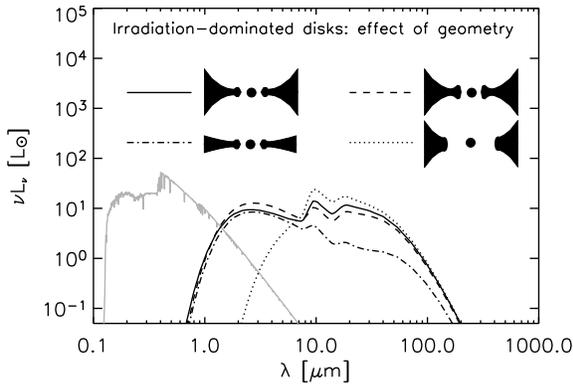}}
\caption{\label{fig-seds-irrad}\revadd{Overall SED shape for {\em
non-accreting} disks with stellar irradiation, computed using
the 2-D radiative transfer tools from {\it Dullemond \& Dominik}
(\citeyear{duldomdisk:2004}). The stellar spectrum is added
in grey-scale. 
Scattered light is not included in these SEDs.
Solid line is normal flaring disk with inner dust rim; dashed line is when
the rim is made higher;
dot-dashed line is when
the flaring is reduced (or when the disk becomes `self-shadowed'); dotted line
is when the inner rim is at 10$\times$ larger radius. 
}}
\end{figure}

\begin{figure}[tb]
\centerline{\includegraphics[width=8cm]{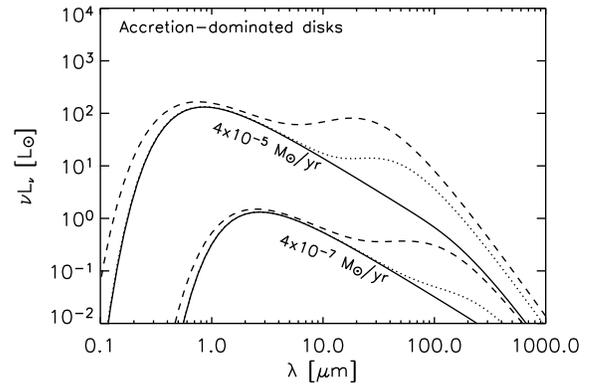}}
\caption{\label{fig-seds-accr}\revadd{Overall SED shape for accreting disks
{\em without} stellar irradiation for two accretion rates. A simple
Shakura-Sunyaev model is used here with grey opacities. Solid line: pure
Shakura-Sunyaev model (star not included); dotted line: model with
disk-self-irradiation included; dashed line: model with
disk-self-irradiation {\em and} irradiation by the magnetospheric accretion
column on the star included. 
}}
\end{figure}

\revised{In Fig.~\ref{fig-seds-irrad} we summarize in a qualitative way how
the disk geometry (inner rim, flaring) affects the SED shape of an
irradiated passive disk. \revadd{In Fig.~\ref{fig-seds-accr} the SEDs of
actively accreting disks are shown, in which the irradiation by the central
star is ignored. In reality, both the accretional heating and the
irradiation by the central star must be included in the models
simultaneously. 
}}

\subsection{2-D radiative transfer in disk models}\label{subsec-2dtrans}
\noindent
The models described so far are all based on an approximate 1+1D (or
two-layer) irradiation-angle description. In reality the structure of these
disks is 2-D, if axisymmetry can be assumed, and 3-D if it cannot. Over the
last 10 years many multi-dimensional dust continuum radiative transfer
programs and algorithms were developed for this purpose (e.g., {\it Whitney
et al.}, \citeyear{whitneyhartmann:1992}; {\it Lucy et al.},
\citeyear{lucy:1999}; {\it Wolf et al.}, \citeyear{wolfhenning:1999}; {\it
Bjorkman and Wood}, \citeyear{bjorkmanwood:2001}; {\it Nicolinni et al.},
\citeyear{niccolini:2003}; {\it Steinacker et al.},
\citeyear{steinacker:2003}). Most applications of these codes assume a given
density distribution and compute spectra and images. There is a vast
literature on such applications which we will not review here (see
\ppvcit{chapter by Watson et al.}). But there is a trend to include the
self-consistent vertical density structure into the models by iterating
between radiative transfer and the vertical pressure balance equation ({\it
Nomura}, \citeyear{nomura:2002}; {\it Dullemond}, \citeyear{dullemond:2002},
henceforth D02; {\it Dullemond and Dominik}, \citeyear{duldomdisk:2004},
henceforth DD04a; {\it Walker et al.},
\citeyear{walkerwood:2004}). \revadd{The main improvements of 2-D/3-D models
over 1+1D models is their ability to account for radial radiative energy
diffusion in the disk, for cooling of the outer disk in radial direction,
for the complex 3-D structure of the dust inner rim, and in general for more
realistic model images.}

\revadd{In addition to this, 2-D/3-D models allow for a `new' class of disk
geometries to be investigated. The 1+1D models can, because of their
reliance on an irradiation angle $\varphi$, only model disk geometries that
are either flat or flared. In principle, however, there might be
circumstances under which, roughly speaking, the surface of the outer disk
regions lies within the shadow of the inner disk regions (although the
concept of `shadow' must be used with care here).
These shadowed regions are cooler than they would be if the disk
was flaring, but the 2-D/3-D nature of radiative transfer prevents them from
becoming entirely cold and flat. For Herbig Ae/Be stars the origin of the
shadow might be the puffed-up inner rim (D02, DD04a), while for T Tauri stars
it might be the entire inner flaring disk region out to some radius
(DD04b).}

\revadd{Although the concept of `self-shadowing' is still under debate, 
it might be linked to
various observable features of protoplanetary disks. For instance, DD04a
showed that self-shadowed disks produce SEDs consistent with Meeus group II
sources, while flaring disks generally produce group I type SEDs, unless the
disk outer radius is very small. It might also underly the observed
correlation between SED shape and sub-millimeter slope ({\it Acke et
al.}~\citeyear{acke:2004a}). Moreover, self-shadowed disks, when spatially
resolved in scattered light, would be much dimmer than flaring disks.}

\section{\textbf{GAS TEMPERATURE AND LINE SPECTRA}}
\label{sec-surface}\noindent
Although \revised{the dust in disks is generally more easily observed}, 
there is an obvious interest in direct
observations of the gas. \revised{It dominates the mass, sets the
structure and impacts dust dynamics and settling in} these disks. Moreover,
it is important to estimate how long disks remain gas-rich, and whether this
is consistent with the formation time scale of gas giant planets ({\it
Hubickyj et al.}, \citeyear{2004RMxAC..22...83H}).  Unfortunately, gas lines
\revised{such as CO rotational, H$_2$ rotational and atomic fine structure
lines} often probe those \revised{surface} regions of disks in which the gas
temperature is difficult to compute. The disk models we described above
assume that the gas temperature in the disk is always equal to the local
dust temperature. While this is presumably true for most of the matter deep
within optically thick disks, in the tenuous surface layers of these disks
(or throughout optically thin disks) the densities become so low that the
gas will thermally decouple from the dust. The gas will acquire its own
temperature, which is set by a balance between various heating- and cooling
processes. These processes depend strongly on the abundance of various
atomic and molecular species, which, for their part, depend strongly on the
temperature. The gas temperature, density, chemistry, radiative transfer and
radiation environment are therefore intimately intertwined and have to be
studied as a \revised{single entity}.  This greatly complicates the modeling
effort, and the first models which study this in detail have only recently
been published.

\revised{This chapter focuses} on stationary models, i.e.~models that are in
chemical, thermal and hydrostatic equilibrium. For the tenuous regions of
disks the chemical time scales are short enough that this is valid, in
contrast to the longer chemical time scales deeper in the disk (e.g.,~{\it
Aikawa and Herbst}, \citeyear{1999A&A...351..233A}; {\it Willacy et
al.}, \citeyear{2000ApJ...544..903W}). The models constructed so far either
solve the gas temperature/chemistry for a {\em fixed} gas density structure
\citep{2004A&A...428..511J, 2004ApJ...615..991K}, or include the gas density
in the computation to obtain a self-consistent thermo-chemical-hydrostatic
structure \citep{2004ApJ...613..424G, 2005A&A...438..923N}.

\subsection{Basic gas physics}\noindent
The physics and chemistry of the surface layers of protoplanetary disks
strongly resembles that of photon dominated regions (PDRs, {\it Tielens and
Hollenbach}, \citeyear{1985ApJ...291..722T}; {\it Yamashita et
al.}~\citeyear{1993ApJ...402L..65Y}). In those \revised{surface layers} the
gas temperature generally greatly exceeds the dust temperature. But the
dust-gas coupling gradually takes over the gas temperature balance as one
gets deeper into the disk, typically beyond a vertical column depth of
\revised{$A_V \simeq 1$}, \revised{and forces the gas temperature to the dust
temperature.}

The uppermost surface layer contains mostly atomic and ionized species,
since the high UV irradiation effectively dissociates all molecules ({\it
Aikawa et al.}, \citeyear{2002A&A...386..622A}). The photochemistry is
driven by the stellar \revised{UV} irradiation and/or in case of nearby O/B
stars, by external illumination. In flaring disk models, the stellar
\revised{UV} radiation penetrates the disk under an irradiation angle 
$\varphi$ like
the one described in the previous section.  This radiation gets diluted with
increasing distance from the central star and attenuated by dust and gas
along an {\em inclined} path into the disk. The stellar \revised{UV}
radiation therefore penetrates less deep into the disk than external UV
radiation. \revised{As one goes deeper into the surface layer, the gas
becomes molecular (see chapter by \ppvcit{Bergin et al.}).}

The thermal balance of the gas in disks is solved by equating all relevant
heating and cooling processes. For this gas thermal balance equation, a
limited set of key atomic and molecular species is sufficient: e.g.,~H$_2$,
CO, OH, H$_2$O, \revised{C$^+$, O, Si$^+$ and various other heavy elements}.
For most atoms and molecules, the statistical equilibrium equation has to
include the pumping of the fine structure and rotational levels by 
the cosmic background radiation, \revised{which
become important deep in the disk, where stellar radiation cannot
penetrate. The full radiative transfer in chemical models is very
challenging, and therefore generally approximated by a simple escape
probability approach, where the pumping and escape probability are derived}
from the optical depth of the line (similar to the approach of {\it Tielens
and Hollenbach}, \citeyear{1985ApJ...291..722T} for PDRs).  \revised{Even
though the emitted photons travel in all directions,} the optical depth used
for this escape probability is the line optical depth in the {\em vertical}
direction where the photons most readily escape.

\revised{One} of the most critical ingredients of these models is the UV and
X-ray radiation field (stellar and external), \revised{which can be split
into the far-ultraviolet (FUV, 6-13.6 eV), the extreme-ultraviolet (EUV,
13.6-100 eV) \ and X-ray ($\ga 100$ eV) regime.} In the literature the far
ultraviolet radiation field \revised{(FUV)} is often represented by a single
parameter $G_0$ describing the integrated intensity between 912 and
2000~\AA\ normalized to that of the typical interstellar radiation field
\citep{1968BAN....19..421H}. However, several papers have shown the
importance of a more detailed description of the radiation field for
calculations of the chemistry and the gas heating/cooling balance
({\it Spaans et al.}, \citeyear{1994ApJ...437..270S}; 
{\it Kamp \& Bertoldi}, \citeyear{2000A&A...353..276K};
{\it Bergin et al.}, \citeyear{2003ApJ...591L.159B}; 
{\it Kamp et al.}, \citeyear{kamp:2005};
\revised{{\it Nomura and Millar}, \citeyear{2005A&A...438..923N}}). For
instance, in T Tauri stars the radiation field is dominated by strong
Ly\,$\alpha$ emission, which has consequences for the photodissociation rate
of molecules that can be dissociated by Ly\,$\alpha$ photons. The
photoelectric heating process, on the other hand, depends strongly on the
overall shape of the radiation field, which is much steeper in the case of
cool stars. A similar problem appears in the X-ray spectra of cool M stars, 
which are dominated by line emission.

\revised{FUV} induced grain photoelectric heating is often a dominant
heating process for the gas in the irradiated surface layers.  \revised{The
FUV photon is absorbed by a dust grain or a polycyclic aromatic hydrocarbon
(PAH) molecule, which ejects an energetic electron to the gas, and heats the
gas via the thermalization of the energetic electron.} Its efficiency and
thus the final gas temperature depends strongly on the grain charge, dust
grain size and composition (PAHs, silicates, graphites, ices, etc.).  X-rays
from the central star \revised{also heat only} the uppermost surface layers,
as \revised{their heating drops off monotonically with column, and gets
quite small by columns of order $10^{21}$~cm$^{-2}$.}
 
\subsection{Surfaces of optically thick disks}\noindent

This subsection focuses on \revised{the warm surface layers} of the optically
thick disk at $A_V < 1$\revised{, measured vertically downwards,} where
\revised{gas and dust temperatures decouple. Modeling of these surface
layers is not affected by the optically thick disk interior and depends
mainly on the local UV/X-ray flux and the gas density. The typical hydrogen
gas number density in these layers is roughly $n_H(A_V=1) \simeq 10^7 \;
\left(100\;\mathrm{AU}/r \right)\;\delta^{-1}\;\mathrm{cm}^{-3}$.  The
location of the $A_V=1$ surface depends on the ratio $\delta$} of the dust
surface area per hydrogen nucleus to the interstellar value, which is
roughly $10^{-21}\mathrm{cm}^2/$H. \revised{Assuming a surface density that
drops linearly with radius and $\Sigma(1~{\rm AU}) = 1000$~g~cm$^{-2}$, the
fractional column density} $\Sigma_{\mathrm{surf}}/\Sigma$ 
contained in the surface layer ($T_{\rm
gas} \neq T_{\rm dust}$) is usually small, $\Sigma_{\mathrm{surf}}/\Sigma
\simeq 1.5\times 10^{-6}\,
\delta^{-1}\, \left(r/{\mathrm{AU}}\right)$, \revised{but increases linearly
with radius.}

\vspace{1em}

\sssec{Gas temperatures.}
\begin{figure}[tb]
\includegraphics[width=8cm]{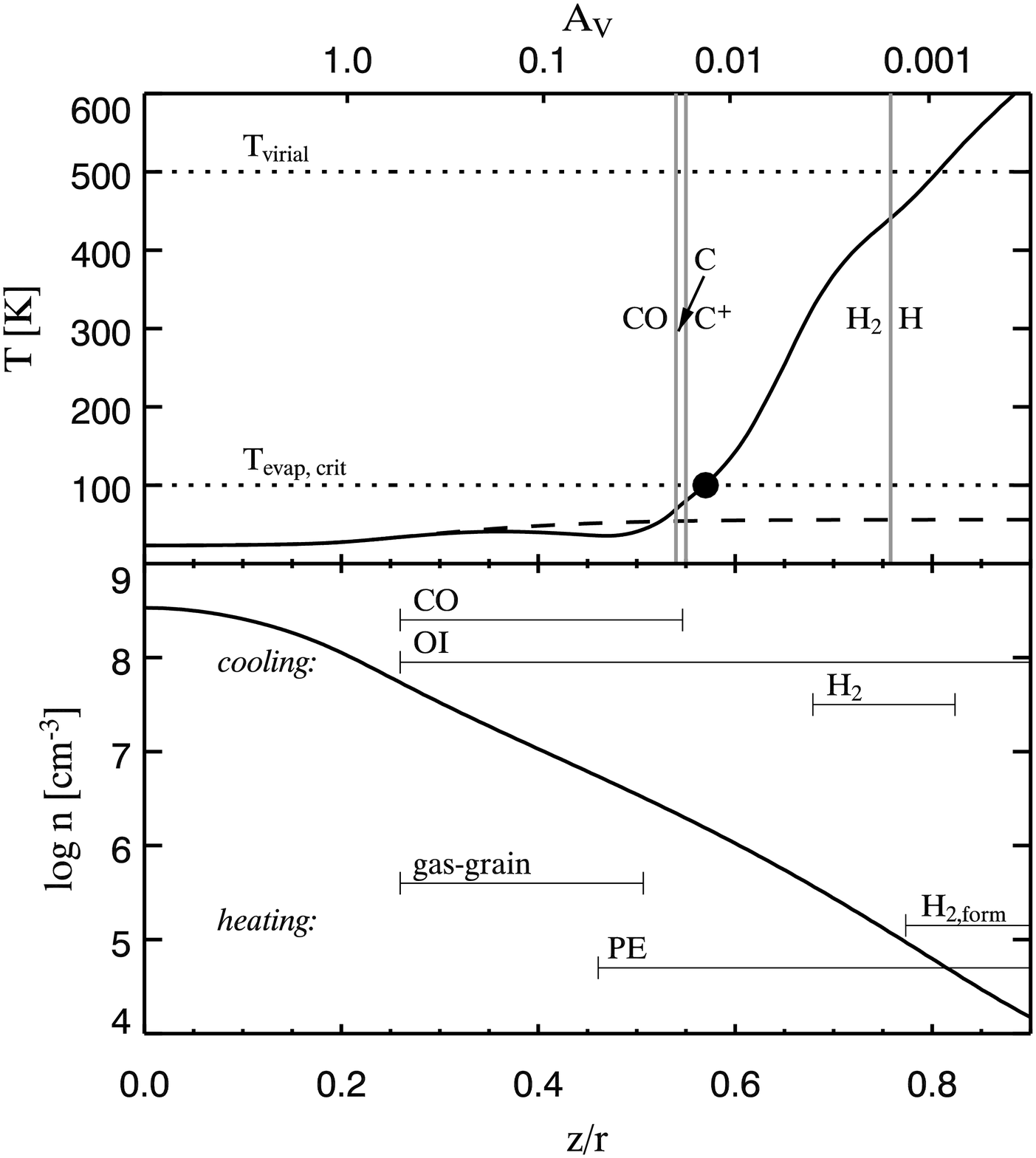}
\caption{\label{100AUslice}
\revised{Upper panel: gas (solid line) and dust (dashed line) temperatures
 in a vertical cut through the T Tauri star model at 100~AU. Overplotted are
 the most important chemical transitions from atomic to molecular
 species. The black filled circle shows the point where a photoevaporative
 flow can be initiated (see Section~\ref{sec-evaporation}). Lower panel:
 same vertical cut for the hydrogen number density (fixed input density
 distribution). Overplotted are the
 ranges over which the respective heating/cooling processes contribute more
 than 10\% to the total heating/cooling rate at that depth (`PE' means
 photoelectric heating; `H$_{2,\mathrm{form}}$' means heating through 
 molecular hydrogen formation; see Section~\ref{sec-evaporation} for the
 definition of $T_{\mathrm{evap,crit}}$ and $T_{\mathrm{virial}}$).}}
\end{figure} 
The detailed temperature structure of the surface layers of optically thick
young disks was studied for the first time by \citet{2004A&A...428..511J},
\citet{2004ApJ...615..991K}, and \citet{2005A&A...438..923N}. \revised{These
models neglect EUV irradiation and start with neutral hydrogen in the top
layers.}  Fig.~\ref{100AUslice} shows the vertical structure in a disk model
with 0.01~$M_\odot$ at 100 AU around a 0.5~$M_\odot$ T Tauri star
\revised{(from models of {\it Kamp and
Dullemond}, \citeyear{2004ApJ...615..991K}; note that the density structure
in these models is not iterated with the gas temperature)}. \revised{Very
high in the atmosphere at particle densities as low as $n < 10^5$~cm$^{-3}$
($A_V \lesssim 10^{-3}$), the gas temperature is set by a balance between
photoelectric heating and} fine structure line cooling of neutral oxygen
\citep{2004ApJ...615..991K, 2004A&A...428..511J}. \revised{This leads to gas
temperatures of several hundred~K. Deeper in the disk, for $A_V > 0.01$,}
molecules can shield themselves from the dissociating \revised{FUV}
radiation.  As soon as the fraction of molecular hydrogen becomes larger
than 1\%, H$_2$ line cooling \revised{becomes important}. Molecular line
emission \revised{-- mainly CO and H$_2$ --} cools the gas down to
\revised{below} hundred K before the densities are high enough for gas and
dust to thermally couple. \revised{As gas
temperatures drop below $\sim 100$~K, H$_2$ no longer contributes} to the
cooling. Instead CO, which has a rich rotational spectrum at low
temperatures, becomes an important coolant.
At larger radii the \revised{FUV flux} from the
central star drops as well as the \revised{density of the surface layer,
leading to lower gas temperatures}. \revised{At distances $r \gtrsim
100$~AU the gas temperature is} too low for the endothermic destruction of
H$_2$ by O atoms and hence the \revised{surface layer at those distances}
contains substantial fractions of molecular hydrogen. 

\vspace{1em}

\sssec{Implications for the disk structure.}
Detailed models of the gas temperature have shown that gas and dust are
collisionally coupled at optical depth $A_V > 1$. Thus the basic assumption
$T_{\rm gas} = T_{\rm dust}$ of the disk structure models presented in the
previous section is justified \revised{for the disk interior. The main
effect of the higher gas temperatures in the warm surface layer is an
enhanced flaring of the disk
} \citep{2005A&A...438..923N}.

\vspace{1em}

\begin{figure}[tb]
\includegraphics[width=8cm]{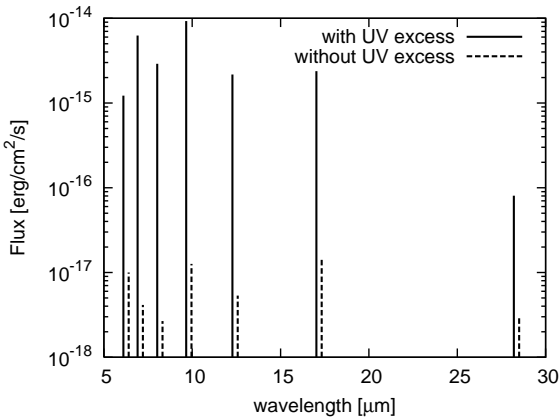}
\caption{\label{H2:nomura}
The mid-infrared line spectra of molecular hydrogen from a T Tauri disk
model ($M_\ast = 0.5$~$M_\odot$, $R_\ast = 2$~$R_\odot$, $T_{\rm
eff}=4000$~K, $\dot M = 10^{-8}$~$M_\odot$/yr) with (solid line) and without
(dotted line) UV excess \revised{({\it Nomura and
Millar}, \citeyear{2005A&A...438..923N})}.}
\end{figure}

\sssec{Observations and comparison with models.} The pure rotational lines of
H$_2$ such as J = 2 -- 0 S(0) [28 $\mu$m], J = 3 -- 1 S(1) [17 $\mu$m], 
J = 4 -- 2 S(2) [12
$\mu$m] and J = 6 -- 4 S(4) [8 $\mu$m] trace the warm gas (100-200 K) in the
disks. \revised{Even though there is some controversy about detection of
those lines with different instruments \citep{2001ApJ...561.1074T,
2002ApJ...572L.161R}, there is a tentative detection of H$_2$ in AB Aurigae
using the Texas Echelon Cross Echelle Spectrograph (TEXES) at the Infrared
Telescope Facility \citep{2002ApJ...572L.161R}. 
\citet{2003ApJ...586.1136B} report \revised{v = 1 -- 0 S(1)
[2.12 $\mu$m]} emission in high resolution spectra ($R\sim 60\,000$) of the
T Tauri stars GG Tau A, LkCa 15, TW Hya and DoAr21. \revised{This emission
most likely arises in the low density, high temperature upper surfaces
beyond 10~AU.} According to the disk models, warm H$_2$ exists indeed in the
optically thin surface layers, where $T_{\rm gas} \gg T_{\rm dust}$ and the
observed fluxes can be reproduced ({\it Nomura and
Millar}, \citeyear{2005A&A...438..923N}). 
Fig.~\ref{H2:nomura} reveals the
effect of UV fluorescence on the line strength.} 
\revised{The UV fluorescent
lines, which are an excellent probe of the inner disk ($r <$ few AU), are
discussed in detail in the \ppvcit{chapter by Najita et al.}}. 
The detection of the
mid-IR H$_2$ lines at low spectral resolution (e.g.,\ with Spitzer) is
hindered by the low line-to-continuum ratio.
 
\begin{figure}[tb]
\includegraphics[width=8cm]{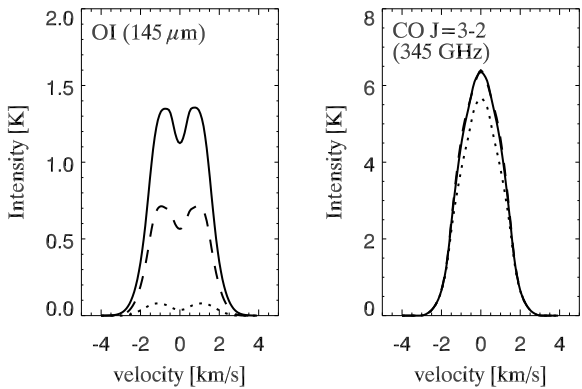}
\caption{\label{jonkheid04}
\revised{Impact of detailed gas temperature modeling and dust settling on
the emission lines of oxygen and CO ({\it Jonkheid et
al.}~\citeyear{2004A&A...428..511J}): $T_{\rm gas} = T_{\rm dust}$ (dotted
line), detailed gas energy balance (solid line), dust settling (dashed
line).}}
\end{figure}

\revised{The impact of detailed gas modeling differs for the various
emission lines. CO, which forms deeper in the disk is generally less
affected than fine structure lines such as [O\,{\sc i}] and [C\,{\sc ii}]
that form in the uppermost surface layers, where $T_{\rm gas} \gg T_{\rm
dust}$ (Fig.~\ref{jonkheid04}). Since the gas temperature in those layers is
set by photoelectric heating, dust settling leads to lower temperatures and
thus to weaker line emission.}

\revised{The [O\,{\sc i}] 6300~\AA\ line is} another tracer of the physics
in the tenuous surface layers (see also the chapter by \ppvcit{Bergin et al.}). 
It has been detected in a number of
externally illuminated proplyds in the Orion nebula
\citep{1998ApJ...499..758J} as well as in T Tauri and Herbig Ae/Be stars
\citep{2005A&A...436..209A,2005astro.ph.12562A}. \cite{1999ApJ...515..669S}
explain \revised{the emission in the Orion proplyds} by the
photodissociation of the OH molecule, which leaves about 50\% of the atomic
oxygen formed in the upper $^1{\rm D}_2$ level of the $6300$~\AA\
line. \cite{2005A&A...436..209A} \revised{find indication of Keplerian
rotation from the [O\,{\sc i}] line profiles. However, they} need OH
abundances higher than those predicted from disk models to fit the emission
from the disks around Herbig Ae/Be stars. \revised{Gas models of those
disks} reveal the presence \revised{of a high temperature reservoir} (few
1000 K); hence the [O\,{\sc i}] line might arise partly from thermal
excitation \revised{at radii smaller than 100~AU ({\it Kamp et
al.}, \citeyear{kamp:2005})}. \revised{Resolved [O\,{\sc i}] 6300~\AA\ line
emission from the disk around the Herbig Ae star HD\,100546 
\citep{2005astro.ph.12562A} shows that the
emission is spread between $\sim 1$ and 100 AU and supports the presence of
a gap at $\sim 10$~AU as reported initially by \citet{bouwmandekoter:2003}.}

\subsection{Optically thin disks}\noindent
\revised{As protoplanetary disks evolve}, the dust grains grow to
\revised{at least} centimeter sizes and the disks become optically thin. In
addition, as we shall discuss \revised{Section~\ref{sec-evaporation}}, the gas
in the disk ultimately disappears, turning the disk into a debris disk. It
is therefore theoretically conceivable that there exists a transition period
in which the disk has become optically thin in dust continuum, but still
contains a detectable amount of gas.  
\revadd{The source HD141569 (5 Myr) might be an example of this, as 
{\it Brittain et al.}~(\citeyear{2003ApJ...588..535B}) observed UV
excited warm CO gas from the inner rim at $\sim$ 17 AU, and {\it Dent et 
al.}~(\citeyear{2005MNRAS.359..663D}) cold gas further out (J = 3 -- 2).}
Measuring the gas
mass in such transition disks sets a timescale for the planet formation
process. The Spitzer Legacy Science Program `Formation and Evolution of
Planetary Systems' (FEPS) has set upper limits on gas masses of $\sim
0.1$~$M_{\rm J}$ around solar-type stars with ages greater than 10~Myr
({\it Meyer et al.}, \citeyear{2004ApJS..154..422M}; 
{\it Hollenbach et al.}, \citeyear{2005ApJ...631.1180H}; {\it Pascucci
et al.}, \citeyear{2005prpl.conf.8468P}).
 
\vspace{1em}
 
\sssec{Disk models.}
Several groups have so far \revised{studied} these transition phases of
protoplanetary disks: \cite{2004ApJ...613..424G} modeled the disk structure
and gas/dust emission from intermediate aged disks around low-mass stars,
\cite{2000A&A...353..276K}, \cite{2001A&A...373..641K}, and
\cite{2003A&A...397.1129K} modeled the gas chemistry and line emission from
A-type stars such as $\beta$~Pictoris and Vega. {\it Jonkheid et al.}
(\citeyear{jonkheid:06}) studied the gas chemical structure and molecular
emission in the \revised{transition phase} disk around HD\,141569\,A. These
models are all based on the same physics as outlined above for the optically
thick protoplanetary disks. The disks are still in hydrostatic equilibrium,
so that the disk structure in these low mass disks is similar to that in the
more massive disks with the midplane simply removed. However, some
fundamental differences remain: the minimum grain size in these disks is
typically a few microns, much larger than in the young protoplanetary disks;
in addition, the dust may have settled towards the midplane, and much of the
solid mass may reside in larger particles ($a > 1$ cm) than can be currently
observed. This reduces the grain opacity and the dust-to-gas mass ratio
compared to the younger optically thick disks.  \revised{At radial midplane
gas column densities smaller than $10^{23}$~cm$^{-2}$, these disks are}
optically thin to stellar UV and $\sim 1$~keV X-ray photons \revised{and the
gas is mostly atomic}. At \revised{radial} columns greater than
\revised{that}, the gas opacity becomes large enough to shield H$_2$ and CO,
allowing significant molecular abundances. For disks extended to 100~AU,
very little mass (very roughly $\gtrsim 10^{-3}$~$M_{\rm J}$) is needed to
provide this shielding.

\vspace{1em}

\begin{figure}[tb]
\includegraphics[width=8cm]{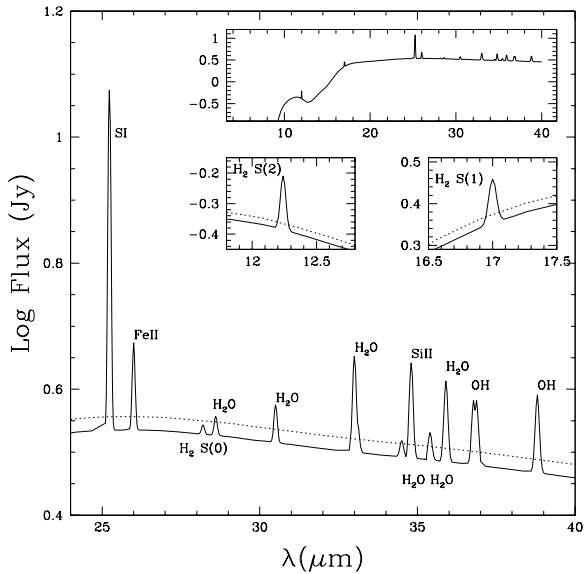}
\caption{\label{GH04:fig8}
\revised{Mid-infrared spectrum in the 24 -- 40~$\mu$m wavelength region
showing the dust continuum and dominant gas emission lines for a disk model
with $M_{\rm gas} = 10^{-2}\, M_{\mathrm{J}}$ and $M_{\rm dust} = 10^{-5}\, 
M_{\mathrm{J}}$ 
(with dust defined as particles smaller than 1 mm). A distance to the
disk of 30 pc and a spectral resolving power $R = 600$ is assumed ({\it
Gorti and Hollenbach}, \citeyear{2004ApJ...613..424G})}}
\end{figure}

\sssec{Comparison with observations.}
\revised{Since the continuum remains optically thin, the mid-IR spectrum is
dominated by fine-structure emission lines from ions such as [Fe\,{\sc ii}]
and [Si\,{\sc ii}]; large columns of neutral sulphur are common, leading to
strong [S\,{\sc i}] emission (Fig.~\ref{GH04:fig8}). However, the strength
and thus detectability of these lines depends on the abundances of
\revised{heavy metals} in late phases of disk evolution, which is uncertain,
especially the more refractory Fe.  In the opaque molecular regions somewhat
closer to the star, the \revised{gas temperature exceeds} 100~K and
molecular hydrogen emission is produced. The S(0) and S(1) H$_2$ lines stay
optically thin over a large range of disk masses and \revised{{\em if
detected} are more diagnostic of disk mass than other fine structure lines}
(\cite{2004ApJ...613..424G}). While the strongest molecular bands of H$_2$O
(important coolant in disk midplane) and OH are similar in strength to the
fine structure lines, the H$_2$ lines are weak and} detection can be
significantly hampered by the low line-to-continuum ratio (weak narrow line
against the bright dust thermal background).  These \revised{mid-IR} lines
generally originate from 1--10~AU.

\cite{2003A&A...397.1129K} have shown that beyond 40~AU the dominant coolant
for the latest tenuous stages of disk evolution is the [C\,{\sc
ii}]~158~$\mu$m line. The fine structure lines of C, O and C$^+$ trace only
the surface of these tenuous disks: \revised{[O\,{\sc i}] becomes rapidly
optically thick and C$^+$ and C turn into CO as soon as UV CO and H$_2$
bands become optically thick, and stellar UV cannot penetrate any further}.
Since typical gas temperatures are higher than in molecular clouds, CO lines
from the upper rotational levels (J = 4 -- 3) are predicted to be stronger
than the lower J lines. \cite{2005MNRAS.359..663D} have recently detected
the CO J = 3 -- 2 line in HD141569 and disk modeling by {\it Jonkheid et
al.}~(\citeyear{jonkheid:06}) shows that the profile excludes a significant
contribution from gas inwards of $\sim 80$~AU and estimate the total gas
mass to \revised{be 80}~$M_{\rm E}$.

\section{\textbf{PHOTOEVAPORATION OF A DISK BY ITS CENTRAL STAR}}
\label{sec-evaporation}

\subsection{Introduction}
%
The above section has shown that in the surface layers of the disk the gas
temperature can become very high, greatly exceeding the dust
temperature. The warm surface gas can flow off the disk and
escape the gravity of the star. Since the heating process responsible
for these high temperatures is the {\em radiation} from the central star or a
nearby O-star, this process is called ``photoevaporation''.  The viscous
evolution (i.e.~accretion and spreading) of the disk, discussed in Section
\ref{sec-viscevol}, can be strongly affected by this photoevaporation
process. Typically, it significantly shortens the `lifetime' of a disk
compared to pure viscous evolution. 
Photoevaporation can also create inner holes or truncate the outer disk.
This has relevance to observations of
such disks, such as the percentage of young stars with infrared excess
versus their age ({\it Haisch et al.}, \citeyear{haischladalada:2001}; {\it
Carpenter et al.}, \citeyear{carpenterwolf:2005}), or the inferred `large
inner holes' of some disks (e.g.,~{\it Calvet et
al.}, \citeyear{calvetdalessio:2002}; {\it Bouwman et
al.}, \citeyear{bouwmandekoter:2003}; {\it Forrest et
al.}, \citeyear{forrest:2004}; {\it D'Alessio et
al.}, \citeyear{2005ApJ...621..461D}). It has also far-reaching consequences
for the formation of planets, as we will discuss below.

Photoevaporation has already been discussed in earlier reviews ({\it
Hollenbach et al.}, \citeyear{2000prpl.conf..401H}; {\it
Hollenbach and Adams}, \citeyear{hollenadams:2004}; {\it Richling et
al.}, \citeyear{richlinghollen:2006}). However, these reviews mainly focused
on the heating by a nearby massive star (such as the famous case of the
proplyds in Orion). In contrast, in this section we will exclusively review
recent results on photoevaporation by the central star, \revised{consistent
with the previous sections which focus on heating and photodissociation by
the central star.}  Progress in this field since PPIV has been mostly
theoretical, since observations of diagnostic gas spectral lines for the
case of photoevaporation by the central, low mass star requires greater
sensitivity, spectral resolution, and spatial resolution than currently
available. We will, however, discuss the implications for the observed
`inner holes' and disk lifetimes.

\subsection{The Physics of Photoevaporation}

\sssec{Basic Concepts.}
Photoevaporation results when stellar radiation heats the disk surface and
resulting thermal pressure gradients drive an expanding hydrodynamical flow
to space. As shown in Section \ref{sec-surface} the main heating photons lie
in the FUV, EUV and X-ray energy regimes. X-rays, however, were shown to
be of lesser importance for photoevaporation ({\it Alexander
et~al.}, \citeyear{2004MNRAS.354...71A}), and we will not consider them
further.

There are two main sources of the strong EUV and FUV excesses observed in
young low mass stars: accretion luminosity and prolonged excess
chromospheric activity.  Recent work ({\it Alexander
et~al.}, \citeyear{2004MNRAS.348..879A}) has shown that EUV photons do not
penetrate accretion columns, so that accretion cannot provide escaping EUV
photons to power photoevaporation.  {\it Alexander
et~al.}~(\citeyear{2005MNRAS.358..283A}) present indirect observational
evidence that an active chromosphere may persist in T Tauri stars even
without strong accretion, and that EUV luminosities of $\Phi _{\mathrm{EUV}} >
10^{41}$ photons/s may persist in low mass stars for extended ($\ga 10^{6.5}
- 10^7$ yrs) periods to illuminate their outer disks.  FUV photons may
penetrate accretion columns and also are produced in active chromospheres.
They are measured in nearby, young, solar mass stars with little accretion
and typically (with great scatter) have luminosity ratios
$L_{\mathrm{FUV}}/L_{\mathrm{bol}}\sim 10^{-3}$ or $\Phi_{\mathrm{FUV}} 
\sim 10^{42}$ photons/s.

EUV photons ionize the hydrogen in the very upper layers of the disk and
heat it to a temperature of $\sim 10^4$ K, independent of radius.  FUV
photons penetrate deeper into the disk and heat the gas to $T\sim 100 -
5000$ K, depending on the intensity of the FUV flux, the gas density and the
chemistry (as was discussed in Section \ref{sec-surface}).  Whether the EUV
or FUV heating is enough to drive an evaporative flow depends on how the
resulting \revised{thermal speed (or sound speed)} compares to the local
escape speed from the gravitationally bound system.  A characteristic radius
for thermal evaporation is the ``gravitational radius'' $r_g$, where the
sound speed equals the escape speed:
\begin{equation}
r_g=\frac{\mbox{G}M_{\star}\mu m_p}{kT}
\sim 100\;\mbox{AU}\;\left(\frac{T}{1000\;\mbox{K}}\right)^{-1}
                     \left(\frac{M_{\star}}{M_{\odot}}\right).
\end{equation}
Early analytic models made the simple assumption that photoevaporation
occurred for $r>r_g$, and that the warm surface was gravitationally bound
for $r<r_g$. However, a closer look at the gas dynamics shows that this
division happens not at $r_g$ but at about $0.1$ -- $0.2\; r_g$ ({\it
Liffman}, \citeyear{liffman:2003}; {\it Adams et
al.}, \citeyear{2004ApJ...611..360A}; {\it Font
et~al.}, \citeyear{2004ApJ...607..890F}), and that this division is not
entirely sharp. In other words, photoevaporation happens {\em mostly}
outside of the ``critical radius'' $r_{\mathrm{cr}} \sim 0.15r_g$, though a
weak evaporation occurs inside of $r_{\mathrm{cr}}$. Since these are
important new insights since PPIV, we devote a subsubsection on them below.

With $T\sim 10^4$ K the critical radius for EUV-induced photoevaporation is
$r_{\mathrm{cr}}(\mathrm{EUV})\sim 1$ -- $2 (M_*/ M_\odot)$ AU.  However, there is
no fixed $r_{\mathrm{cr}}(\mathrm{FUV})$ because the FUV-heated gas has temperatures
that depend on FUV flux and gas density, i.e., on $r$ and $z$. Therefore,
$r_{\mathrm{cr}}(\mathrm{FUV})$ depends on $r$ and $z$, and may range from 3-150 AU
for solar mass stars.

The evaporative mass flux $\dot\Sigma$ depends not only on the temperature
of the photon-heated gas, but also on the vertical penetration depth of the
FUV/EUV photons. For EUV photons this is roughly set \revised{for $r<r_{\mathrm{cr}}
\sim 1$ AU} by the Str\"omgren condition that recombinations in the ionized
layer equal the incident ionizing flux. Neglecting dust attenuation, this
penetration column \revised{can be expressed: $A_V(\mathrm{EUV})\sim 0.05 
\,\delta\, \Phi_{41}^{1/2} (r/\mathrm{AU})^{-1/2}$, where 
$\Phi_{41}\equiv\Phi_{\mathrm{EUV}}/10^{41}$ photons/s 
and $\delta$ is the ratio of the dust
surface area per hydrogen to the interstellar dust value (see Section 
4 and note
that $\delta$ can be much smaller than unity if dust has settled or
coagulated). Outside of 1 AU, the penetration depth falls even faster with
$r$, roughly as $r^{-3/2}$ (see Hollenbach et al. 1994).  On the other hand,
the FUV penetration depth is set by dust attenuation, or 
$A_V(\mathrm{FUV}) \sim
\varphi$, where we recall that $\varphi$ is the irradiation angle and
depends on disk flaring.  In general $A_V(\mathrm{EUV})\ll A_V(\mathrm{FUV})$, so the
EUV-ionized skin of the disk lies on top of the FUV-heated gas surface
layer.}

The penetration depth is an important quantity because it sets the density
at the base of the photoevaporative flow: the deeper the penetration depth,
the higher the density. The flux of outflowing matter is
\revised{proportional to} the product of local density and sound speed
within this heated layer. This is why the complex surface structure models
of Section \ref{sec-surface} are so important for FUV-driven
photoevaporation. For EUV-driven photoevaporation, on the other hand, the
situation is less complicated, since the temperature in the ionized skin of
the disk is independent of $r$ and $z$, as long as $z>z_b$, where $z_b$ is
the bottom of the ionized layer, i.e.~the base of the flow. For this simple
case, the evaporative mass flux \revised{ originates at $z_b$, which is
where the highest density gas at temperature $T_{\mathrm{EUV}}\simeq 
10^4\,$K resides}.

Although FUV-heated layers have lower temperatures than the
EUV-heated skin they are at higher densities and may equally well initiate
the flow and determine the mass flux as EUV photons (see {\it Johnstone 
et~al.}, \citeyear{1998ApJ...499..758J} for a similar situation for
externally illuminated disks). \revised{{\it Gorti and Hollenbach} (in
preparation, henceforth GH06) find that the FUV-photoevaporative flow
typically originates at vertical heights where $T \sim 100-200$ K, yielding
$r_{\mathrm{cr}} \sim 50-100$ AU. For $r>50$ AU, the FUV photoevaporation
dominates.  On the other hand, EUV photons (with $r_{\mathrm{cr}}\sim 1$ AU) affect 
the planet forming regions at
$r\ll 50$ AU more than the FUV photons.}

\vspace{1em}

\sssec{Photoevaporation as a Bernoulli flow.} One way to understand why the
disk can evaporate at radii as small as $0.2r_g$ is to consider the
evaporative flow as a Bernoulli flow along streamlines ({\it
Liffman}, \citeyear{liffman:2003}; {\it Adams et
al.}, \citeyear{2004ApJ...611..360A}). These streamlines initially rise
nearly vertically out of the disk and then bend over to become
asymptotically radially outward streamlines.  \revised{If a streamline
starts at $r>r_g$, then the flow rapidly goes through a sonic point and
achieves the sound speed $c_s$ near the base of the flow.  The mass flux
rate in the flow is then $\dot \Sigma \simeq \rho _b c_s$, where $\rho _b$
is the mass density of the gas at the base.}

\revised{On the other hand, if a streamline starts at $r\ll r_g$,} the gas at
its base lies deep in the gravitational potential. As a simplification let
us now treat these streamlines as if they are entirely radial streamlines
(ignoring their vertical rise out of a disk).  Then the standard atmospheric
solution has a density that falls off from $r$ to roughly $r_g$ as
exp$(-r_g/2r)$.  The gas flows subsonically and accelerates, as it slowly
expands outward, until it passes through a sonic point at $r_s\la 0.5 r_g$
($0.5r_g$ is the classic Parker wind solution for zero rotation).  For
$r\ll r_g$, the mass flux is reduced considerably by the rapid fall-off of the
density from $r$ to $r_s$.  For $r < r_g$, the mass flux is roughly given by
the density at $r_s$ times the sound speed times the dilution factor
$(r_s/r)^2$ that accounts for mass conservation between $r$ and $r_s$: $\dot
\Sigma \simeq \rho_b e^{-r_g/2r}c_s (r_s/r)^2$.  Assuming the same $\rho_b$
and $c_s$ at all $r$, we see that $\dot \Sigma (0.2r_g) \simeq 0.5\dot
\Sigma (r_g)$ and that $\dot \Sigma (0.1r_g) \simeq 0.17 \dot \Sigma (r_g)$.
This demonstrates that $r_{\mathrm{cr}} \sim 0.15 r_g$ for this simplified
case, and that even for $r\lesssim r_{\mathrm{cr}}$ evaporation is weak, but
not zero. In Fig.~\ref{100AUslice} the base of the flow is marked with the
large dot (though that figure shows a static, non-evaporating model with
only FUV heating). In
that figure, $T_{\mathrm{virial}}$ is the temperature such that the sound
speed equals the escape speed; $T_{\mathrm{evap,crit}} \equiv
0.2T_{\mathrm{virial}}$ is roughly where the photoevaporation flow
originates (i.e., where $r=r_{\mathrm{cr}}$).

\vspace{1em}

\sssec{Mass loss rates for EUV-induced flows.} Although central star FUV
models are not yet published, several central star EUV models have appeared
in the literature. {\it Hollenbach et~al.}~(\citeyear{1994ApJ...428..654H})
first outlined the essential physics of EUV-induced flows by the central star
and presented an approximate analytic solution to the mass loss rate for a
disk larger than $r_g$.  The basic physics is the Str\"omgren relation, $\Phi
_{\mathrm{EUV}} \simeq \alpha _r n_e^2 r^3$, where $\alpha _r$ is the hydrogen
recombination coefficient and $n_e$ is the electron density in the ionized
surface gas.  This sets the hydrogen nucleus (proton) number density at the
base of the flow $n_b \propto \Phi _{\mathrm{EUV}}^{1/2}$, and therefore an identical
proportionality for the mass loss rate:
 \begin{equation}\label{equ_mdot_central}
\dot{M}_{\rm EUV}\sim4\times10^{-10}
     \left(\frac{\Phi_{\rm EUV}}{10^{41}\;\mbox{s}^{-1}}\right)^{0.5}
     \left(\frac{M_{\star}}{M_{\odot}}\right)^{0.5}
\end{equation}
in units of $M_{\odot}/\mbox{yr}$.
Radiation hydrodynamical simulations ({\it Yorke and
Welz}, \citeyear{1996A&A...315..555Y}; {\it Richling and
Yorke}, \citeyear{1997A&A...327..317R}) find a similar power-law index for
the dependence of the mass-loss rate on the EUV photon rate of the central
star.  This result applies for both high and low mass central stars, and is
valid for a weak stellar wind.  The effect of a strong stellar wind is such
that the ram pressure reduces the scale height of the atmosphere above the
disk and the EUV photons are allowed to penetrate more easily to larger
radii. This increases the mass-loss rate from the outer parts of the disk.
It is noteworthy that the diffuse EUV field, caused by recombining electrons
and protons in the disk's ionized atmosphere inside $r_{\mathrm{cr}}$,
controls the EUV-induced mass-loss rates ({\it Hollenbach
et~al.}, \citeyear{1994ApJ...428..654H}) for disks \revised{with no or small
inner holes ($<r_{\mathrm{cr}}$)}.  This effect negates any potential for
self-shadowing of the EUV by the disk.

\subsection{Evolution of photoevaporating disks}

\sssec{Case without viscous evolution.}
Let us first assume a disk that does not viscously evolve: it just passively
undergoes photoevaporation.  For disks with size $r_d <
r_{\mathrm{cr}}$, the photoevaporation proceeds from outside in.  The mass
flux rate at $r_d$ is much higher than inside of $r_d$, because the gas at
$r_d$ is least bound.  In addition, disk surface densities generally fall
with $r$ (see Section \ref{sec-viscevol}). 
Therefore, the disk shrinks as it photoevaporates, and most of
the mass flux comes from the outer disk radius.  However, for disks with
$r_d>r_{\mathrm{cr}}$, two types of disk evolution may occur.  For EUV
photons, {\it Hollenbach et~al.}~(\citeyear{1994ApJ...428..654H}) showed
that the mass flux $\dot \Sigma$ beyond $r_{\mathrm{cr}}$ goes roughly as
$r^{-2.5}$ \revised{if there is no inner hole extending to
$r_{\mathrm{cr}}$.} The timescale for complete evaporation at $r$ goes as
$\Sigma (r)/ \dot \Sigma (r)$. As long as $\Sigma$ does not drop faster than
$r^{-2.5}$, the disk will evaporate first at $r\sim r_{\mathrm{cr}}$, and,
once a gap forms there, will then steadily erode the disk from this gap
outwards.

If, on the other hand, $\Sigma (r)/ \dot \Sigma (r)$ decreases with $r$,
then the disk shrinks from outside in as in the $r_d < r_{\mathrm{cr}}$
case. The photoevaporation by the FUV from the central star has not yet been
fully explored, but preliminary work by GH06 suggests that the mass flux
$\dot \Sigma$ in the outer disks around solar mass stars {\it increases}
with $r$.  In this case, the disk evaporates from outside in for most
generally assumed surface density laws, which decrease with
$r$. \revised{The combined effect of EUV and FUV photoevaporation then is
likely to erode the disk outwards from 
$r_{\mathrm{cr}}(\mathrm{EUV})\sim 1$ AU by
the EUV flow and inwards from the outer disk radius by the FUV flow,
sandwiching the intermediate radii.}

\vspace{1em}

\sssec{Case with viscous evolution.}
%
Now let us consider a disk that is actively accreting onto the star (see
Section \ref{sec-viscevol}). In general, if the photoevaporation drills a
hole somewhere in the disk or `eats' its way from outside in, the forces of
viscous spreading tend to move matter toward these photoevaporation regions,
which can accelerate the dissipation of the disk. If the disk has a steady
accretion rate $\dot M$, then a gap forms once $\dot M_{\mathrm{evap}} \propto r^2
\dot \Sigma$ exceeds $\dot M$.  Since $r^2 \dot \Sigma \propto r^{-0.5}$
\revised{for EUV photoevaporation beyond $r_{\mathrm{cr}}$}, the gap first forms at
the minimum radius ($\sim r_{\mathrm{cr}}$) and then works its way
outward. {\it Clarke et~al.}~(\citeyear{2001MNRAS.328..485C}) presented
time-dependent computations of the evolution of disks around low mass stars
with $\Phi_{\mathrm{EUV}}\sim 10^{41-43}$ photons s$^{-1}$. Their model combines EUV
photoevaporation with a viscous evolution code. \revised{ However, they used
the current hypothesis at that time that evaporation only occurs outside of
$r_g$.} After $\sim 10^6$ to 10$^7$ years of viscous evolution relatively
unperturbed by photoevaporation, the viscous accretion inflow rates fall
below the photoevaporation rates at $r_g$. At this point, a gap opens up at
$r_g$ and the inner disk rapidly (on an inner disk viscous timescale of
$\sim 10^5$ yr) drains onto the central star or spreads to $r_g$ where it
evaporates. In this fashion, an inner hole is rapidly produced extending to
$r_g$.

\begin{figure}[tb]
\centerline{\includegraphics[width=8cm]{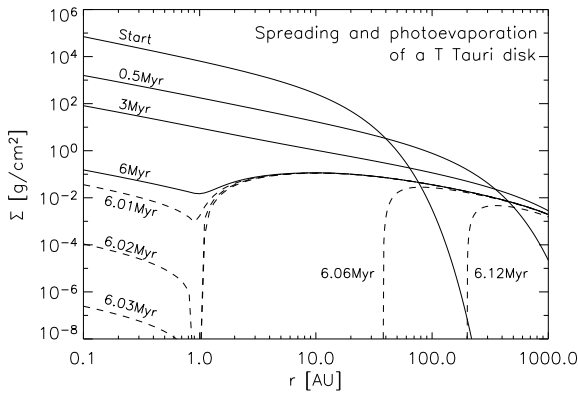}}
\caption{\label{fig-alexander-sigma}
Evolution of the surface density of a EUV-photoevaporating disk (Figure
adapted from {\it Alexander et al.,}~\citeyear{alexander:06b}).  This
simulation starts from a given disk structure of about 0.05 $M_{\odot}$
(marked with `Start' in the figure). Initially the disk accretes and
viscously spreads (solid lines).  At $t=6\times 10^6\,$yr the photoevaporation
starts affecting the disk. Once the EUV-photoevaporation has drilled a gap
in the disk at $\sim 1$ AU, the destruction of the disk goes very rapidly
(dashed lines).The inner disk quickly accretes onto the star, followed by a
rapid erosion of the outer disk from inside out. In this model the disk
viscosity spreads to $> 1000$ AU; however, FUV-photoevaporation (not included)
will likely truncate the outer disk.}
\end{figure} 

{\it Alexander et al.}~(\citeyear{alexander:06a}, \citeyear{alexander:06b})
extended the work of {\it Clarke et al.} \revised{to include the effect of 
$r_{\mathrm{cr}} < 
r_g$, and to treat the outward EUV evaporation of the disk beyond $r_{\mathrm{cr}}
\sim 1$ AU.}  They show that once the inner hole is produced, the diffuse
flux from the atmosphere of the inner disk is removed and the attenuation of
the direct flux by this same atmosphere is also removed.  This enhances the
EUV photoevaporation rate by the direct EUV flux from the star, and the
effect magnifies as the inner hole grows as $\dot M_{\mathrm{EUV}} \propto
r_{\mathrm{inner}}^{1/2}$, again derivable from a simple Str\"omgren criterion.
The conclusion is that the
outer disk is very rapidly cleared once the inner hole forms \revadd{(see
Fig.~\ref{fig-alexander-sigma})}.

\begin{figure}[tb]
\centerline{\includegraphics[width=7cm]{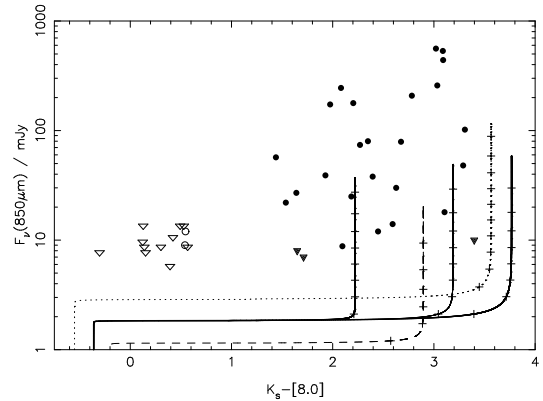}}
\caption{\label{fig-alexander}
Near- to mid-infrared color (in magnitudes) versus 850 $\mu$m flux for
photoevaporation/viscous evolution models.  The data are taken from {\it 
Hartmann et al.}~(\citeyear{hartmann:2005}) and 
{\it Andrews and Williams} (\citeyear{andrewswilliams:2005}): 
850 $\mu$m detections (circles)
and upper limits (triangles) are plotted for both CTTSs (filled symbols) and
WTTSs (open symbols).  Evolutionary tracks are shown for models with stellar
masses 0.5 (dashed), 1.0 (solid), and 2.0 $M_{\odot}$(dotted), at a disk
inclination of $i=60^{\circ}$ to the line of sight.  
The thick tracks to the right and
left show the 1 $M_{\odot}$ model at $i=0$ and $i= 80^{\circ}$, respectively.
Crosses are added every 1 Myr to show the temporal evolution.  Initially the
(optically thin) 850 $\mu$m flux declines slowly at constant (optically
thick) infrared color.  However, 
once the viscous accretion rate falls below the photoevaporation rate,
the disk is rapidly cleared from the inside-out.  (Figure
adapted from {\it Alexander et al.}~\citeyear{alexander:06b}.)
}
\end{figure}

The rapid formation of a cleared out inner hole almost instantly changes the
nature and appearance of the disk. The above authors \revised{compare their
model favorably with a number of observations: (i) the rapid transition from
classical T~Tauri stars to weak line T~Tauri stars, (ii) the almost
simultaneous loss of the outer disk (as detected by submillimeter
measurements of the dust continuum) with the inner disk (as detected by near
IR observations of very hot dust near the star), and (iii) the SED
observations of large (3--10 AU) inner holes in those sources \revadd{(see
dotted line of Fig.~\ref{fig-seds-irrad})} with evidence for low accretion
rates and intermediate mass outer disks 
such as the source CoKu Tau/4.  Fig.~\ref{fig-alexander} shows the
evolutionary tracks of their models with $\Phi _{\mathrm{EUV}} = 10^{42}$ photons/s
compared to the observations of weak-line T Tauri stars (WTTSs) and CTTSs.  }

\revised{In a similar vein {\it Armitage et
al.}~(\citeyear{2003MNRAS.342.1139A}) used the combination of EUV
photoevaporation and viscous dispersal, together with an assumed dispersion
of a factor of 3 in the initial disk mass, to explain the dispersion in the
lifetime and accretion rates of T Tauri disks.  They found that the models
predict a low fraction of binaries that pair a classical T Tauri star with a
weak-lined T Tauri star.  Their models are in better agreement with
observations of disk lifetimes in binaries than models without
photoevaporation.}

\revised{Going one step further, {\it Takeuchi et
al.}~(\citeyear{2005ApJ...627..286T}) constructed models combining viscous
evolution, EUV photoevaporation, and the differential radial motion of
grains and gas.  Their models predicted that for low-mass stars with a low
photoevaporation rate, dust-poor gas disks with an inner hole would form
(WTTs), whereas for high mass stars (evolved Herbig Ae/Be) with a high
photoevaporation rate, gas-poor dust rings would form.}

{\it Matsuyama et~al.}~(\citeyear{2003ApJ...582..893M}) pointed out that if
the EUV luminosity is created by accretion onto the star, then, as the
accretion rate diminishes, the EUV luminosity drops and the timescale to
create a gap greatly increases. Even worse, as discussed above, the EUV
photons are unlikely to escape the accretion column.  Only if the EUV
luminosity remains high due to chromospheric activity does EUV
photoevaporation play an important role in the evolution of disks around
isolated low mass stars.  {\it Alexander et
al.}~(\citeyear{2005MNRAS.358..283A}) argue this is the case.  {\it
Ruden}~(\citeyear{2004ApJ...605..880R}) provides a detailed analytic
analysis which describes the evolution of disks in the presence of viscous
accretion and photoevaporation and compares his results favorably with these
two groups.

\subsection{Effect on planet formation}

The processes which disperse the gas influence the formation of planets.
Small dust particles follow the gas flow.  If the gas is dispersed before
the dust can grow, all the dust will be lost in the gas dispersal and
planetesimals and planets will not form.  Even if there is time for
particles to coagulate and build sufficiently large rocky cores that can
accrete gas ({\it Pollack et~al.}, \citeyear{1996Icar..124...62P}; {\it
Hubickyj et~al.}, \citeyear{2004RMxAC..22...83H}), the formation of gas
giant planets like Jupiter and Saturn will be suppressed if the gas is
dispersed before the accretion can occur. Furthermore, gas dispersal helps
trigger gravitational instabilities that may lead to planetesimal formation
({\it Goldreich and Ward}, \citeyear{goldward:1973}; {\it Youdin and
Shu}, \citeyear{youdinshu:2002}; {\it Throop
et~al.}, \citeyear{2005ApJ...623L.149T}), affects planet migration
(e.g.,~{\it Ward}, \citeyear{1997Icar..126..261W}) and influences the orbital
parameters of planetesimals and planets ({\it Kominami and
Ida}, \citeyear{2002Icar..157...43K}).

\vspace{1em}

\sssec{Gas Rich Versus Gas Poor Giant Planets in the Solar System.} {\it Shu
et~al.}~(\citeyear{1993Icar..106...92S}) showed that with $\Phi_{\mathrm{EUV}}\sim
10^{41}$ photons s$^{-1}$, the early Sun could have photoevaporated the gas
beyond Saturn before the cores of Neptune and Uranus formed, leaving them
gas poor. However, this model ignored photoevaporation inside of $r_g$. The
current work by {\it Adams et~al.}~(\citeyear{2004ApJ...611..360A}) would
suggest rather rapid photoevaporation inside of 10 AU, and make the timing
of this scenario less plausible.  FUV photoevaporation (either from external
sources or from the central star) may provide a better explanation.
Preliminary results from GH06 suggest that the early Sun did not produce
enough FUV generally to rapidly remove the gas in the outer giant planet
regions.  {\it Adams et~al.} and {\it Hollenbach and Adams},
(\citeyear{hollenadams:2005}) discuss the external illumination case,
\revised{which looks more plausible}.

\vspace{1em}

\sssec{Truncation of the Kuiper Belt.} A number of observations point to the
truncation of Kuiper Belt Objects (KBOs) beyond about 50 AU (e.g., {\it
Allen, Bernstein, and Malhotra}, \citeyear{2002AJ....124.2949A}; {\it
Trujillo and Brown}, \citeyear{2001ApJ...554L..95T}).  {\it Adams} {\em
et~al.}~(\citeyear{2004ApJ...611..360A}) and {\it Hollenbach and Adams}
(\citeyear{hollenadams:2004}, \citeyear{hollenadams:2005}) show that
photoevaporation by a nearby massive star could cause truncation of KBOs at
about 100 AU, but probably not 50 AU. The truncation is caused by the gas
dispersal before the dust can coagulate to sizes which survive the gas
dispersal, and which can then later form KBOs.  Models of FUV
photoevaporation by the early Sun are needed.

\vspace{1em}

\sssec{Formation of Planetesimals.} In young disks, dust settles toward the
midplane under the influence of the stellar gravity and coagulates.  Once
coagulated dust has concentrated in the midplane, the roughly
centimeter-sized particles can grow further by collisions or by local
gravitational instability ({\it Goldreich and
Ward}, \citeyear{goldward:1973}; {\it Youdin and
Shu}, \citeyear{youdinshu:2002}).  A numerical model by {\it Throop and
Bally}~(\citeyear{2005ApJ...623L.149T}) follows the evolution of gas and
dust independently and considers the effects of vertical sedimentation and
external photoevaporation. The surface layer of the disk becomes
dust-depleted which leads to dust-depleted evaporating flows.  Because of
the combined effects of the dust settling and the gas evaporating, the
dust-to-gas ratio in the disk midplane is so high that it meets the
gravitational instability criteria of {\it Youdin and
Shu}~(\citeyear{youdinshu:2002}), indicating that kilometer-sized
planetesimals could spontaneously form. These results imply that
photoevaporation may even trigger the formation of planetesimals.
Presumably, photoevaporation by the central star may also produce this
effect.

\section{\textbf{SUMMARY AND OUTLOOK}}
\label{sec-outlook}
\begin{figure*}
\mbox{}\vspace{1em}\\
\parbox[t]{8.2cm}{
\centerline{Dust-structure of disk}
\centerline{\parbox[b]{7.5cm}{\mbox{}\vspace{1.4cm}\\
\includegraphics[width=7.5cm]{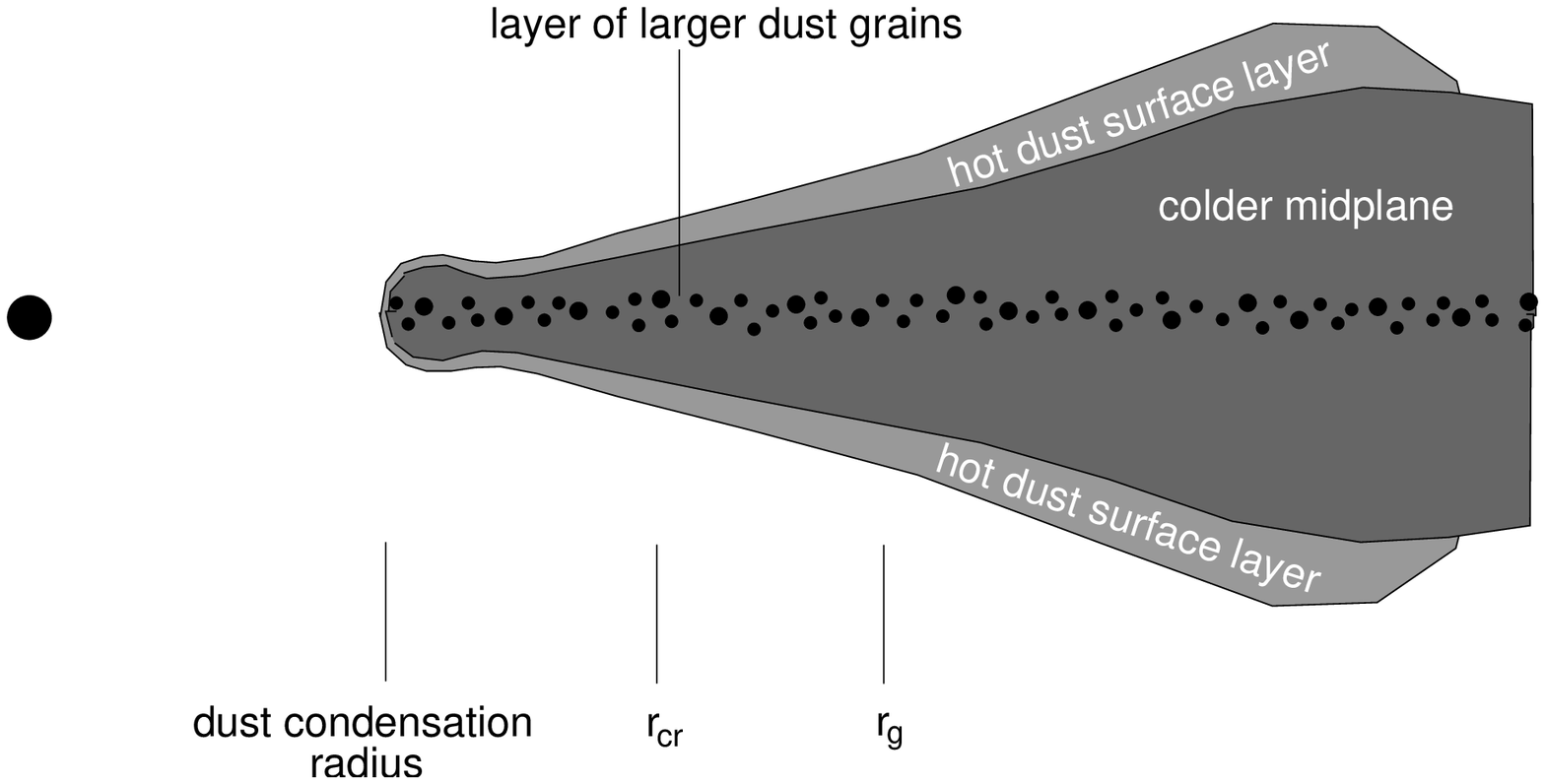}
\\\vspace{1.2cm}\mbox{}}}
}
\parbox[t]{8.2cm}{
\centerline{Gas structure of disk}
\centerline{\includegraphics[width=7.5cm]{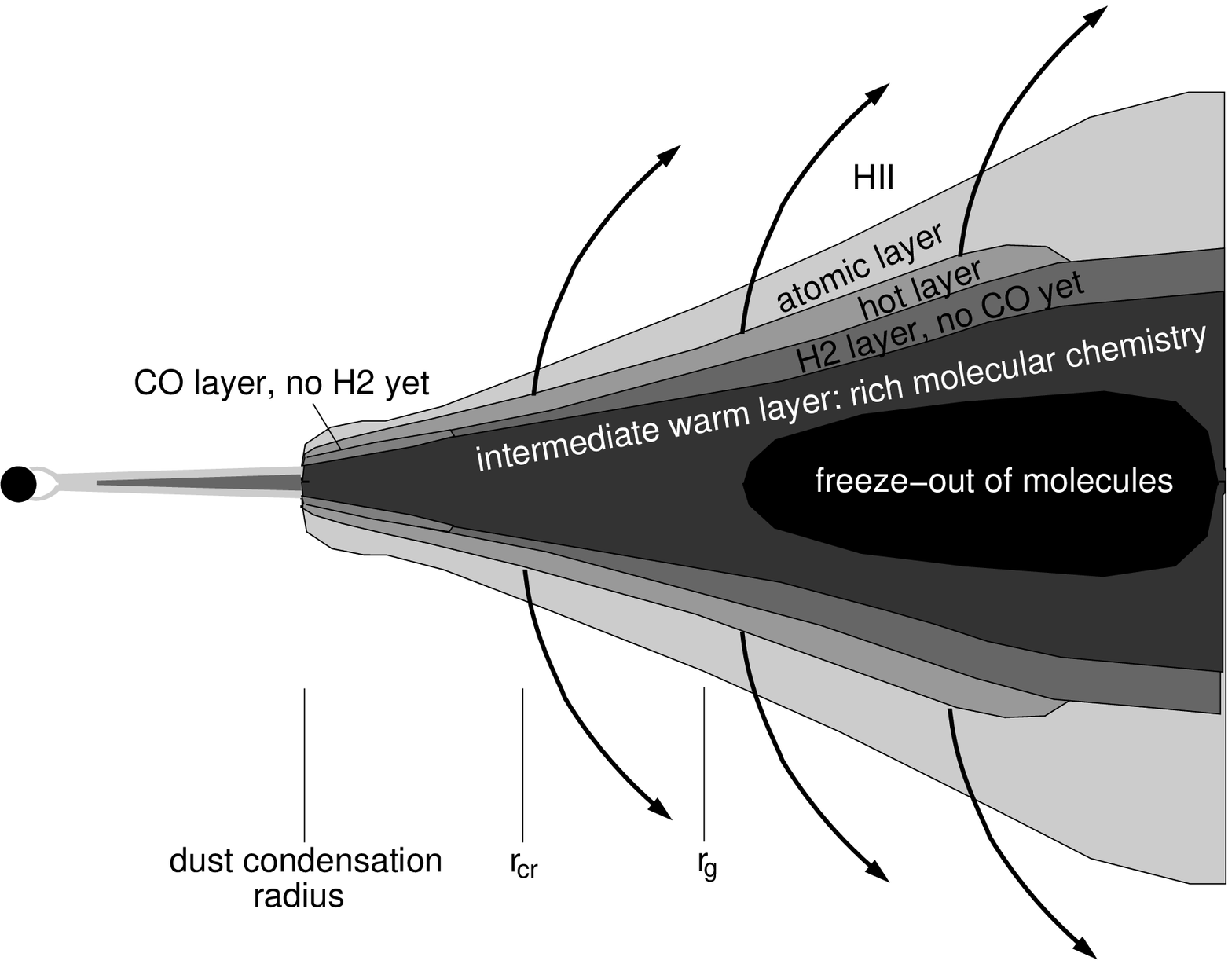}}
}
\caption{\label{fig-picto}Pictograms of the structure of a flaring
protoplanetary disk, in dust (left) and gas (right).}
\end{figure*}

In this chapter we have given a brief outline of how disks form and
viscously evolve, what their structure is, what their spectra look like in
dust continuum and in gas lines, and how they might come to their end by
photoevaporation and viscous dispersion.
The disk structure in dust and gas
is summarized in Fig.~\ref{fig-picto}. Evidently, due to the broadness of
the topic we had to omit many important issues. For instance the formation
of disks is presumably much more chaotic than the simple picture we have
discussed. In recent years there is a trend to outfit even the
observation-oriented workhorse models with ever more detailed physics. This
is not a luxury, since the volume of observational data (both spectral and
spatial) is increasing dramatically, as shown by various other chapters in
this book. For instance, with observational information about dust growth
and sedimentation in disks, it will be necessary to include realistic dust
evolution models into the disk models. Additionally, with clear evidence for
non-axial symmetry in many disks (e.g.,~{\it Fukagawa et
al.}, \citeyear{fukagawa:2004}) modelers may be forced to abandon the
assumption of axial symmetry. The thermal-chemical study of the gas in the
disk surface layers is a rather new field, and more major developements are
expected in the next few years, both in theory and in the comparison to
observations. These new insights will also influence the models of
FUV-photoevaporation, and thereby the expected disk lifetime.

A crucial step to aim for in the future is the unification of the various
aspects discussed here. They are all intimitely connected together and
mutually influence each other.
Such a unification opens up the
perspective of connecting seemingly unrelated observations and thereby
improving our understanding of the bigger picture.

\vspace{1\baselineskip}

\textbf{ Acknowledgments:} 
The authors would like to thank Al Glassgold, Malcolm Walmsley, Klaus
Pontoppidan and Jes Joergensen for proof-reading and providing very useful
comments.

\bigskip


\end{document}